\documentclass[prb,twocolumn,superscriptaddress,showpacs]{revtex4}

\usepackage[dvips]{graphics}
\usepackage{amssymb}
\usepackage{amsmath}
\usepackage{epsfig}
\usepackage{latexsym}
\usepackage{color}
\usepackage{rotating}
\usepackage{float}

\begin{document}

\newcommand{\boxedeqn}[1]{%
  \[\fbox{%
      \addtolength{\linewidth}{-2\fboxsep}%
     \addtolength{\linewidth}{-2\fboxrule}%
      \begin{minipage}{\linewidth}%
     \begin{equation}#1\end{equation} %
     \end{minipage}%
   }\]%
}

\title{Quartz-superconductor quantum electromechanical system}
\author{M. J. Woolley} 
\affiliation{School of Engineering and IT, UNSW Canberra, ACT, 2600, Australia}
\author{M. F. Emzir}
\affiliation{School of Engineering and IT, UNSW Canberra, ACT, 2600, Australia}
\author{G. J. Milburn }
\affiliation{ARC Centre for Engineered Quantum Systems, School of Mathematics and Physics, University of Queensland, St Lucia, 4072, Australia}
\author{M. Jerger }
\affiliation{ARC Centre for Engineered Quantum Systems, School of Mathematics and Physics, University of Queensland, St Lucia, 4072, Australia}
\author{M. Goryachev }
\affiliation{ARC Centre for Engineered Quantum Systems, School of Physics, University of Western Australia, Perth, 6009, Australia}
\author{M. E. Tobar }
\affiliation{ARC Centre for Engineered Quantum Systems, School of Physics, University of Western Australia, Perth, 6009, Australia}
\author{A. Fedorov }
\affiliation{ARC Centre for Engineered Quantum Systems, School of Mathematics and Physics, University of Queensland, St Lucia, 4072, Australia}

\begin{abstract}
We propose and analyse a quantum electromechanical system composed of a monolithic quartz bulk acoustic wave (BAW) oscillator coupled to a superconducting transmon qubit via an intermediate \emph{LC} electrical circuit. Monolithic quartz oscillators offer unprecedentedly high effective masses and quality factors for the investigation of mechanical oscillators in the quantum regime. Ground-state cooling of such mechanical modes via resonant piezoelectric coupling to an \emph{LC} circuit, which is itself sideband cooled via coupling to a transmon qubit, is shown to be feasible. The fluorescence spectrum of the qubit, containing motional sideband contributions due to the couplings to the oscillator modes, is obtained and the imprint of the electromechanical steady-state on the spectrum is determined. This allows the qubit to function both as a cooling resource for, and transducer of, the mechanical oscillator. The results described are relevant to any hybrid quantum system composed of a qubit coupled to two (coupled or uncoupled) thermal oscillator modes. 
\end{abstract}


\maketitle

\section{Introduction}
Recent experiments have demonstrated the cooling of macroscopic mechanical oscillators to their quantum ground state \cite{cleland,teufel:cool,painter}, as well as the generation of quantum squeezed states \cite{schwab:squeezed,sillanpaa,teufel:squeeze}. This work provides a foundation for the demonstration of entangled quantum states of mechanical modes \cite{woolley:entangled} and enhanced sensing capabilities \cite{woolley:twomode}. In most of the experimental demonstrations \cite{teufel:cool,painter,schwab:squeezed,sillanpaa,teufel:squeeze}, the mechanical motion of membranes and beams modulates parameters of a high-frequency electromagnetic cavity mode, forming a cavity optomechanical system \cite{milburnwoolley,aspelmeyerreview}. Driving of the cavity mode enables cooling of the mechanical oscillator, analogous to the laser cooling of trapped ions to their motional ground state \cite{leibfried}. However, parametric coupling of this type is ineffective for quartz bulk acoustic wave (BAW) oscillators. 

Fortunately, quartz oscillators can be directly coupled to electrical circuits due to the piezoelectric effect \cite{gautschi}. Resonant coupling of a \emph{film} BAW oscillator to a superconducting qubit has been realised \cite{cleland}, and indeed, provided the first observation of a macroscopic mechanical degree of freedom in its quantum ground state. However, unlike film BAW oscillators, monolithic BAW oscillators offer exceptional mechanical properties including large effective masses and extremely high quality factors \cite{rayleigh,BAWgeometry}. This makes them an attractive platform not only for the pursuit of quantum optics experiments with phonons, but also for tests of the limits of quantum mechanics itself \cite{pikovski,marin}, high-frequency gravitational wave detection \cite{BAWGWD}, and tests of Lorentz symmetry \cite{lo}. 

On the other hand, their large geometric size makes coupling to them challenging. Most critically, large unavoidable stray capacitance between the BAW oscillator electrodes reduces the amplitude of the oscillating voltage, and the corresponding coupling strength to electrical circuits becomes impractically small. To maximise the coupling between an electrical circuit and the mechanics we propose a scheme where the stray capacitance itself forms an $LC$ electrical circuit with an additional external shunting inductor. Tuning the $LC$ circuit into resonance with a particular mechanical mode of the BAW oscillator allows for the direct coupling of phonons to photons with greater coupling strength than is possible via the more conventional detuned capacitive coupling schemes \cite{woolley:nanomech}. Now the $LC$ electrical circuit will not be at a sufficiently high frequency to be in its quantum ground state, even in a cryogenic environment. Thus, the $LC$ circuit itself must be cooled: this may be achieved via sideband coupling \cite{blais:sideband,beaudoin,strand} to a superconducting transmon qubit \cite{koch,houck}. The latter forms a circuit QED system \cite{wallraff,blais}, albeit one in which the circuit is at a much lower frequency than the transmon \cite{ilichev,grajcar}. This infrastructure also provides the hardware for quantum state control beyond the ground state.

Note that the cooling and measurement of a macroscopic mechanical oscillator via \emph{direct} coupling to a quantum two-level system has been studied both theoretically \cite{wilsonrae,rabl,martin,zhang,hauss,evers} and experimentally \cite{lahaye,arcizet,lukin,richard,pigeau,ovartchaiyapong,barfuss,reserbat}. However, in our case the direct coupling between a quartz oscillator and a transmon is too weak \cite{lahaye}, and hence effective coupling is not feasible without an intermediate $LC$ tank circuit. 

The study of hybrid quantum systems composed of a solid-state quantum two-level system, an electrical circuit mode, and a mechanical oscillator mode, has attracted considerable interest recently. In theoretical work, Restrepo \emph{et al.} have solved the corresponding \emph{Hamiltonian} problem (with the full radiation pressure interaction) in terms of qubit-cavity-mechanical polaritons \cite{restrepo}. Accounting for dissipative dynamics, they have also described the possibility of cooling and unconventional phonon statistics. Note that this solution is inapplicable here since in our case the \emph{bare} electromechanical coupling Hamiltonian is quadratic, and therefore the qubit-circuit polariton number does not commute with the total Hamiltonian. Others have discussed state engineering possibilities enabled by a cavity-mediated interaction between a qubit and a mechanical oscillator \cite{pflanzer} and by three-body interactions \cite{abdi}.

In terms of experimental work, Pirkkalainen \emph{et al.} have coupled a microwave cavity to a mechanical oscillator via a qubit \cite{pirkk}. Differently from our proposal, the qubit is used as a mechanism for coupling to the mechanics rather than as an auxiliary cooling system. They did, however, observe motional sidebands in this work. They subsequently used the intermediate qubit to greatly enhance the effective optomechanical coupling \cite{pirkkalainen}. Lecocq \emph{et al.} have used a phase qubit to control a mechanical oscillator, with the interaction mediated via a microwave electrical circuit \cite{lecocq}. Here, time-dependent control was used for the measurement of the mechanical oscillator. The key difference from our proposal, aside from the absence of a quartz oscillator, is that in this work the microwave electrical circuit is itself at a relatively high frequency, being near-resonant with the qubit and at a far higher frequency than the mechanical oscillator. The optomechanical interaction in their case is the driven, linearised optomechanical interaction. As noted, such coupling is difficult for quartz BAW oscillators.    

Here we propose and thoroughly analyse a quantum electromechanical system composed of a quartz BAW oscillator coupled to a transmon via an intermediate $LC$ electrical circuit. In Sec.~II we give an overview of BAW oscillator technology, provide an equivalent electrical circuit, and use it to obtain a Hamiltonian description of the system. In Sec.~III we determine the steady-state of the system in both an adiabatic limit and a sideband picture, demonstrating the feasibility of ground-state cooling. In Sec.~IV, we calculate the qubit fluorescence spectrum analytically in the adiabatic limit and numerically in a sideband picture. We demonstrate the existence of motional sidebands, potentially enabling transduction of the mechanical motion. 

\section{System}

\subsection{BAW Oscillators}
The mechanical part of our system is provided by a BAW oscillator. BAW oscillators, originally developed in the frequency control community, can be divided into three main groups: High-Overtone Bulk Acoustic Resonators (HBAR) \cite{Baron:2013aa}, Thin-Film Bulk Acoustic Resonators (FBAR) \cite{Krishnaswamy:4aa} and single-crystal (monolithic) BAW oscillators \cite{jvig}. The latter group is mainly composed of bulk \emph{quartz} devices, which can be used to achieve the highest frequency stability in the RF band. For these devices, an acoustic wave is excited in the thickness of a single crystal plate clamped from the sides via electrodes. 

As noted above, a film BAW oscillator (FBAR) has already been measured in its quantum ground state \cite{cleland}. Their very high resonance frequencies mean that they can be prepared in their quantum ground state with high fidelity in a cryogenic environment. Further, they can be integrated on a chip due to their small size and low mass. On the other hand, FBARs have relatively low quality factors. In contrast, single-crystal BAW oscillators \cite{Galliou2011,Goryachev2012,rayleigh} have losses limited only by material properties with logarithmic temperature dependencies leading to extraordinary quality factors at cryogenic temperatures \cite{Galliou2013}. 

In particular, we consider BVA-type phonon trapping single-crystal BAW oscillators \cite{Besson1977}. Such an oscillator is a thin plate with one curved surface allowing effective trapping of acoustic phonons in the plate centre \cite{BAWgeometry}. This trapping results in the spatial separation of the vibrating parts of the plate from points of suspension, and consequently to the material acoustic loss limit. Several modes of vibration are possible, and each mode of vibration gives a series of overtones corresponding to a different number of acoustic half-waves in the device thickness. We could model many mechanical modes via a parallel connection of $RLC$ branches as per the well-known Butterworth-Van Dyke model \cite{BVD}. However, since the quality factors and resonance frequencies are high, the modes are well-resolved in frequency space and we are justified in considering the coupling to one mechanical mode alone. 

\subsection{Hamiltonian}
\begin{figure}[h]
\begin{center}
$\begin{array}{c}
	\includegraphics[width=0.45\textwidth]{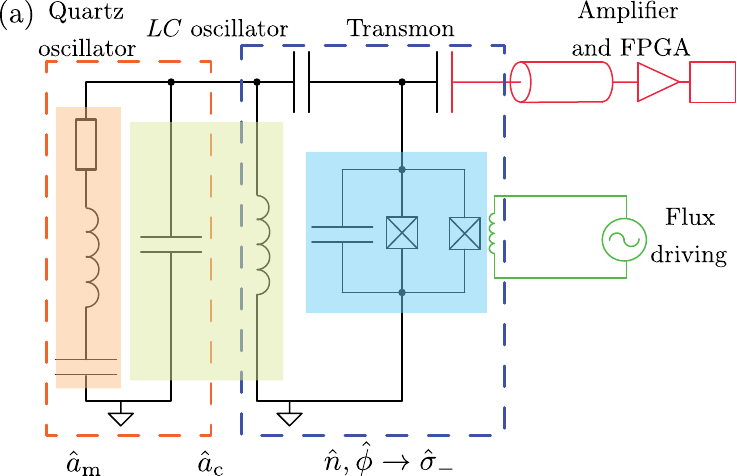} \\ 
	\includegraphics[width=0.45\textwidth]{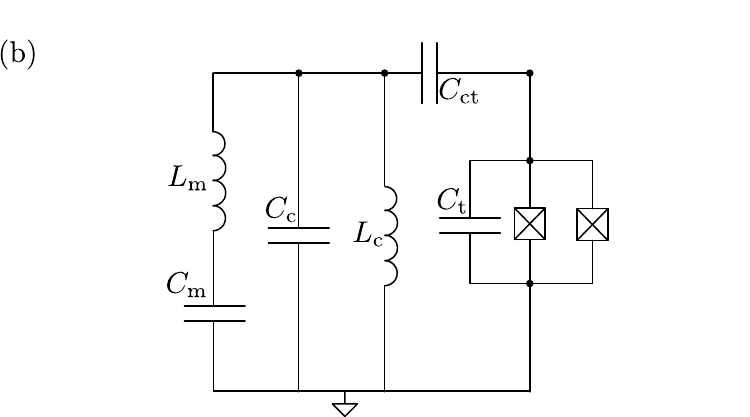}
\end{array}$
\end{center}
	\caption{ (a) Schematic of the system under consideration. The system is composed of a BAW quartz oscillator (orange shading), with electrodes placed across it forming the capacitance of an $LC$ tank circuit (green shading). The quartz (mechanical) oscillator is represented by an electrical equivalent circuit \cite{gautschi}. The $LC$ tank circuit is completed by a tunable inductance on a superconducting chip, which also contains a superconducting transmon qubit (blue shading). This is represented by two Josephson junctions in parallel with a shunt capacitor. Physically, the elements of the BAW oscillator and superconducting chip are demarcated by the dashed orange and blue boxes, respectively. The transmon qubit is flux-driven (green circuit), and the qubit's fluorescence is monitored by capacitive out-coupling to a waveguide and amplification. (b) Equivalent (dissipation-free) electrical circuit for the whole quartz-superconductor quantum electromechanical system. In labelling the components, the subscripts ``m'', ``c'' and ``t'' denote the \emph{mechanical}, electrical \emph{circuit}, and \emph{transmon} modes, respectively. Such an equivalent circuit enables the derivation of the Hamiltonian (\ref{eq:firstHam}), as described in App.~A. } 
	\label{fig:SystemSchematic}
\end{figure}

Now we formulate a minimal model of the proposed system for the purpose of analysis. The system consists of a mechanical oscillator, in the form of a quartz oscillator, coupled to a superconducting $LC$ tank circuit, which is itself sideband coupled to a superconducting circuit in the form of a transmon. The coupling between the mechanical oscillator and the $LC$ circuit is due to the piezoelectric nature of quartz. The inductor of the $LC$ circuit may be realised using a series of DC SQUIDs \cite{wahlsten,Manucharyan2009}, which can be tuned via an external magnetic field to match the resonance of the desired overtone of the quartz oscillator. The system is represented schematically in Fig.~1(a). There is also a small direct coupling between the mechanical oscillator and the transmon. We may write down an equivalent electrical circuit for this electromechanical system \cite{koch}, including an equivalent electrical representation of the mechanical oscillator \cite{gautschi,BVD}. 

The equivalent (dissipation-free) electrical circuit is shown in Fig.~1(b). It may be quantised in the standard manner \cite{devoret:fluctuations}, as described in App.~A, and the resulting Hamiltonian is
\begin{eqnarray}
\hat{H} & = & \hbar \sum_{\rm p} \omega_{\rm p} \hat{a}^\dagger_{\rm p} \hat{a}_{\rm p} + 4 E_{\rm C} \hat{n}^2 + E_{\rm J} (1 - \cos \hat{\phi} ) \nonumber \\
& & + \hbar g_{\rm mc} (\hat{a}_{\rm m} + \hat{a}^\dagger_{\rm m}) (\hat{a}_{\rm c} + \hat{a}^\dagger_{\rm c}) \nonumber \\
& & + 2e \sum_{\rm p} V^{\rm p}_0 \beta_{\rm pt} (\hat{a}_{\rm p} + \hat{a}^\dagger_{\rm p}) \hat{n} , \label{eq:firstHam}
\end{eqnarray}
where the index ${\rm p}$ is summed over the set $\left\{{\rm m},{\rm c} \right\}$, denoting the \emph{mechanical} mode and the electrical \emph{circuit} mode (i.e., the $LC$ tank circuit), respectively. Throughout this Article, the index ${\rm p}$ shall be used in this way. The transmon mode is described by the observables $\hat{n}$ (the number of Cooper pairs transferred between the superconducting islands of the transmon) and $\hat{\phi}$ (the phase difference between the islands). The resonance frequencies of the mechanical oscillator and electrical circuit oscillator are given (in terms of equivalent electrical circuit parameters) by $\omega^2_{\rm m} = 1/L_{\rm m} \tilde{C}_{\rm m}$ and $\omega^2_{\rm c} = 1/L_{\rm c} \tilde{C}_{\rm c}$, respectively. The transmon charging energy is $E_{\rm C}=e^2/2C_{\Sigma}$ and the Josephson energy is $E_{\rm J}=E_{\rm J,max} | \cos (\pi \Phi/\Phi_0 ) |$ where $\Phi$ is the applied magnetic flux and $\Phi_0 = h/(2e)$ is the magnetic flux quantum. In the transmon coupling terms in Eq.~(\ref{eq:firstHam}), $V^{\rm p}_0 = ( \hbar \omega_{\rm p}/2\tilde{C}_{\rm p})^{1/2}$ denotes the rms ground-state voltage fluctuations of the equivalent electrical circuit modes. The effective capacitances, electromechanical coupling ($g_{\rm mc}$) and oscillator-transmon couplings ($\beta_{\rm pt}$) are complicated functions of the equivalent circuit capacitances and inductances. They are fully specified in App.~A.   

We truncate the transmon mode to its two lowest-lying energy levels and use Pauli operators defined in the uncoupled eigenbasis of the resulting qubit. Assuming qubit driving of amplitude $\mathcal{E}_{\rm d}$ and frequency $\omega_{\rm d}$, implemented via modulation of the flux bias \cite{strand}, the Hamiltonian (\ref{eq:firstHam}) takes the form
\begin{subequations}
\begin{eqnarray}
\hat{H}_{\rm S} & = & \hat{H}^{\rm mc}_{\rm S} + \hat{H}^{\rm q}_{\rm S} + \sum_{\rm p} \hat{H}^{\rm pq}_{\rm S} , \label{eq:SchrodingerHam} \\
\hat{H}^{\rm mc}_{\rm S} & = & \hbar \sum_{\rm p} \omega_{\rm p} \hat{a}^\dagger_{\rm p} \hat{a}_{\rm p} + \hbar g_{\rm mc} (\hat{a}_{\rm m} +\hat{a}^\dagger_{\rm m} ) (\hat{a}_{\rm c}+\hat{a}^\dagger_{\rm c} ) , \nonumber \\
& & \\
\hat{H}^{\rm q}_{\rm S} & = & \hbar (\Omega /2) \hat{\sigma}_z - \hbar (\mathcal{E}_{\rm d}/2) \cos \omega_{\rm d} t \, \hat{\sigma}_z , \\
\hat{H}^{\rm pq}_{\rm S} & = & \hbar g_{\rm pq} (\hat{a}_{\rm p}+\hat{a}^\dagger_{\rm p} ) \hat{\sigma}_x , \label{eq:2d}
\end{eqnarray}
\end{subequations}
where the level splitting of the transmon qubit is 
\begin{equation}
\hbar \Omega = \sqrt{8E_{\rm C} E_{\rm J}}-E_{\rm C} , 
\end{equation}
and the couplings of the qubit to the oscillators are given by
\begin{equation}
\hbar g_{\rm pq} = e V^{\rm p}_0 \beta_{\rm pt} (E_{\rm J}/2 E_{\rm C})^{1/4} . \label{eq:qubitcouplings}
\end{equation}
The ${\rm S}$ subscript in Eq.~(\ref{eq:SchrodingerHam}) indicates that the Hamiltonian is specified in a Schr\"{o}dinger picture. The ${\rm q}$ subscript in Eq.~(\ref{eq:qubitcouplings}) indicates that the transmon is being approximated as a qubit. 

For the purpose of determining the steady-state of the system, it is useful to remove the trivial time-dependences in the Hamiltonian (\ref{eq:SchrodingerHam}) and retain only near-resonant couplings between the oscillators and the qubit. To do so, we apply the unitary transformation 
\begin{eqnarray}
\hat{U} & = & \exp \, [ i\sum_{\rm p}\omega_{\rm p} \hat{a}^\dagger_{\rm p} \hat{a}_{\rm p} t + i (\omega_{\rm d}/2) \hat{\sigma}_z t  \nonumber \\
& & - i \mathcal{E}_{\rm d}/(2 \omega_{\rm d}) \sin \omega_{\rm d} t \, \hat{\sigma}_z ] ,
\end{eqnarray}
to (\ref{eq:SchrodingerHam}) under the assumption that the qubit frequency is much higher than the oscillator frequencies (i.e., $\Omega , \omega_{\rm d} \gg \omega_{\rm p}$). This leaves the interaction picture Hamiltonian (I subscript),
\begin{subequations}
\begin{eqnarray}
\hat{H}_{\rm I} & = &  \hat{H}^{\rm q}_{\rm I} + \hat{H}^{\rm mc}_{\rm I} + \sum_{\rm p } \hat{H}^{\rm pq}_{\rm I} , \label{eq:timedependentHam} \\
\hat{H}^{\rm q}_{\rm I} & = & \hbar ( \delta_{\rm d} /2 ) \hat{\sigma}_z , \label{eq:justqubitbit} \\
\hat{H}^{\rm mc}_{\rm I} & = & \hbar g_{\rm mc} (\hat{a}_{\rm m} e^{-i\omega_{\rm m} t} + \hat{a}^\dagger_{\rm m} e^{+i\omega_{\rm m} t} ) \nonumber \\
& & \times (\hat{a}_{\rm c} e^{-i\omega_{\rm c} t} + \hat{a}^\dagger_{\rm c} e^{+i\omega_{\rm c} t} ) \nonumber \\
& \approx & \hbar g_{\rm mc} ( \hat{a}^\dagger_{\rm m} \hat{a}_{\rm c} + \hat{a}_{\rm m} \hat{a}^\dagger_{\rm c} ) , \label{eq:HmclRWA} \\
\hat{H}^{\rm pq}_{\rm I} & = & \hbar g_{\rm pq} (\hat{a}_{\rm p} e^{-i\omega_{\rm p} t} + \hat{a}^\dagger_{\rm p} e^{+i\omega_{\rm p} t}) \nonumber \\
& & \times \sum^{+\infty}_{n=-\infty} \left[ J_{-n}(\mathcal{E}_{\rm d}/\omega_{\rm d} ) e^{i (n+1) \omega_{\rm d} t} \hat{\sigma}_+ \right. \nonumber \\
& & \ \ \ \ \ \ \left. + J_{n}(\mathcal{E}_{\rm d}/\omega_{\rm d} ) e^{i(n-1)\omega_{\rm d} t} \hat{\sigma}_- \right] \nonumber \\
& \approx & \hbar \bar{g}_{\rm pq} (\hat{a}_{\rm p} e^{-i\omega_{\rm p} t} + \hat{a}^\dagger_{\rm p} e^{+i\omega_{\rm p} t}) \hat{\sigma}_x , \label{eq:dcqubitcoupling}
\end{eqnarray}
\end{subequations}
where we have assumed that the mechanical oscillator mode is resonant with the electrical circuit mode $(\omega_{\rm m}=\omega_{\rm c})$ and made a rotating-wave approximation on the electromechanical coupling in (\ref{eq:HmclRWA}). Further, we have introduced $\delta_{\rm d} = \Omega - \omega_{\rm d}$ as the detuning between the qubit level splitting and the qubit drive frequency, and 
\begin{equation}
\bar{g}_{\rm pq} = g_{\rm pq} J_{1} (\mathcal{E}_{\rm d}/\omega_{\rm d} ) ,
\end{equation}
as the \emph{sideband-reduced} couplings in Eq.~(\ref{eq:dcqubitcoupling}), where $J_m(z)$ denotes a Bessel function of the first kind.

\subsection{Dissipation}
The mechanical oscillator and electrical circuit modes $\hat{a}_{\rm p}$ are assumed to be linearly damped at rates $\gamma_{\rm p}$ into independent Markovian environments with thermal occupations $\bar{n}_{\rm p} \equiv n(\omega_{\rm p})$. The qubit is assumed to be damped into a Markovian environment with relaxation, excitation and dephasing rates being denoted by $\gamma_\downarrow$, $\gamma_\uparrow$, and $\gamma_\phi$, respectively. Therefore, the master equation describing the evolution of the system density matrix is 
\begin{subequations}
\begin{eqnarray}
\dot{\rho} & = & -\frac{i}{\hbar} [ \hat{H}, \rho ] + \mathcal{L}^{\rm d}_{\rm mc} \rho + \mathcal{L}^{\rm d}_{\rm q} \rho = \mathcal{L} \rho , \label{eq:ME} \\ 
\mathcal{L}^{\rm d}_{\rm mc} \rho & = & \sum_{\rm p} \left[ \gamma_{\rm p} (\bar{n}_{\rm p}+1) \mathcal{D}[\hat{a}_{\rm p}] \rho + \gamma_{\rm p} \bar{n}_{\rm p} \mathcal{D}[\hat{a}^\dagger_{\rm p} ] \rho \right] , \label{eq:oscillatordissipation} \\ 
\mathcal{L}^{\rm d}_{\rm q} \rho & = & \gamma_\downarrow \mathcal{D}[\hat{\sigma}_-]\rho + \gamma_\uparrow \mathcal{D}[\hat{\sigma}_+] \rho - ( \gamma_\phi /4) \left[ \hat{\sigma}_z , \left[ \hat{\sigma}_z , \rho \right] \right] ,\nonumber \\
& & \label{eq:qubitdissipation}
\end{eqnarray}
\end{subequations}
where $\hat{H}$ is given by (\ref{eq:timedependentHam}), we have introduced the notation $\mathcal{L}$ for the Liouvillian of the entire system, and $\mathcal{L}^{\rm d}_{\rm mc}$ and $\mathcal{L}^{\rm d}_{\rm q}$ for the Lindbladians (i.e., dissipation) of the oscillator modes and qubit, respectively. As is usual, $\mathcal{D}[\hat{c}]$ denotes the dissipative superoperator whose action is given by 
\begin{equation}
\mathcal{D}[\hat{c}] \rho = \hat{c}\rho \hat{c}^\dagger - \frac{1}{2}\hat{c}^\dagger \hat{c} \rho - \frac{1}{2}\rho \hat{c}^\dagger \hat{c}.     
\end{equation}

\subsection{Parameters}
\label{sec:anticipatedparameters}
In order to proceed with the analysis, let us first consider the parameters expected for the proposed system. The parameters appearing in the Hamiltonian~(\ref{eq:SchrodingerHam})-(\ref{eq:2d}) and in the master equation (\ref{eq:ME}) are quoted here; the equivalent electrical circuit parameters from which they are derived are quoted in App.~\ref{app:EquivalentCircuit}. The mechanical and electrical circuit resonance frequencies, and qubit level splitting, are $\omega_{\rm m}/2\pi = 250 \, {\rm MHz}$, $\omega_{\rm c}/2\pi = 250 \, {\rm MHz}$, and $\Omega/2\pi = 8 \, {\rm GHz}$, respectively. The direct mechanics-circuit, circuit-qubit, and mechanics-qubit couplings are $g_{\rm mc}/2\pi = 7 \, {\rm kHz}$, $g_{\rm cq}/2\pi = 20 \, {\rm MHz}$, and $g_{\rm mq}/2\pi = 1 \, {\rm kHz}$, respectively. 

The qubit driving conditions are set by $\omega_{\rm d}/2\pi = (\Omega - \omega_{\rm c})/2\pi = 7.75 \, {\rm GHz}$ and $\mathcal{E}_{\rm d}/2\pi = 775 \, {\rm MHz}$, such that $J_1(\mathcal{E}_{\rm d}/\omega_{\rm d}) = 0.05$.
The effective (sideband-reduced) coupling rates are then $\bar{g}_{\rm cq}/2\pi = 1 \, {\rm MHz}$ and $\bar{g}_{\rm mq}/2\pi = 50 \, {\rm Hz}$. 

The mechanical and electrical circuit mode damping rates are $\gamma_{\rm m}/2\pi = 0.1 \, {\rm Hz}$ and $\gamma_{\rm c}/2\pi = 100 \, {\rm kHz}$, respectively, and the corresponding environmental thermal occupations are $\bar{n}_{\rm m} \sim \bar{n}_{\rm c} = 1.21$ (assuming a cryogenic environment, with $T=20 \, {\rm mK}$). The qubit relaxation, excitation and pure dephasing are given by $\gamma_\downarrow/2\pi = 10 \, {\rm MHz}$, $\gamma_\uparrow/2\pi = 10 \, {\rm kHz}$, and $\gamma_{\phi}/2\pi = 10 \, {\rm kHz}$, respectively.

We note that the parameters specified place us well into the \emph{resolved-sideband} regime \cite{milburnwoolley,aspelmeyerreview,leibfried}, here defined by the condition: 
\begin{equation}
\omega_{\rm c}, \omega_{\rm m} \gg \gamma_{\rm t} , \label{eq:resolvedsideband}
\end{equation}
where $\gamma_{\rm t}$ is the \emph{total} qubit decoherence rate, given by
\begin{equation}
\gamma_{\rm t} = \gamma_\downarrow + \gamma_\uparrow + 2\gamma_\phi . \label{eq:totalqubitdecoherence}
\end{equation}
The parameters specified are also such that we are in an \emph{adiabatic} regime \cite{gardiner} in which the qubit is damped rapidly compared with other relevant time-scales in the system. More precisely, this is here defined by the condition: 
\begin{equation}
\gamma_\downarrow \gg g_{\rm mc}, \bar{g}_{\rm pq}, \gamma_{\rm p} . \label{eq:adiabaticlimit}
\end{equation}
The system shall subsequently be analysed in both of these regimes. 

\section{Steady-state}
\label{sec:steadystate}

The qubit may be used to cool both the mechanical and electrical circuit modes. This may be efficiently achieved in the resolved-sideband regime and the adiabatic limit. Analysis of the system is then facilitated by the adiabatic elimination of the qubit \cite{cirac:ion}, which can (after some further approximations) result in a linear, time-invariant, Markovian description of the dynamics of the reduced system (i.e., the mechanical oscillator and the electrical circuit mode). The analytical adiabatic limit results shall be validated using numerical results obtained in a sideband picture.  

\subsection{Adiabatic Limit}

Adiabatic elimination in the presence of a time-dependent coupling, as in the Hamiltonian~(\ref{eq:dcqubitcoupling}), may be treated using a projection operator approach \cite{gardiner,cirac:ion}. The calculation is detailed in App.~\ref{app:AdiabaticTimeDependent}. We obtain a master equation in Lindblad form for the reduced density matrix of the two oscillator modes, $\rho_{\rm s} = {\rm Tr}_{\rm q}[ \rho ]$,  
\begin{eqnarray}
\dot{\rho}_{\rm s} & = & - \frac{i}{\hbar } [ \hat{H}^{\rm mc}_{\rm I} , \rho_{\rm s} ] + \mathcal{L}^{\rm d}_{\rm mc} \rho_{\rm s} + \sum_{ {\rm p} } ( - i\delta_{\rm p} [\hat{a}^\dagger_{\rm p} \hat{a}_{\rm p} ,\rho_{\rm s} ] \nonumber \\
& & + \gamma^-_{\rm pe} \mathcal{D}[\hat{a}_{\rm p}] \rho_{\rm s} + \gamma^+_{\rm pe} \mathcal{D}[\hat{a}^\dagger_{\rm p} ] \rho_{\rm s}  \nonumber \\
& &  + \bar{g}_{\rm pq} \bar{g}_{\rm \bar{p}q} \{ G(+\omega_{\rm p}) [\hat{a}_{\rm \bar{p}} \rho_{\rm s}, \hat{a}^\dagger_{\rm p} ] \nonumber \\
& & + G(-\omega_{\rm p}) [ \hat{a}^\dagger_{\rm \bar{p}} \rho_{\rm s}, \hat{a}_{\rm p} ] + {\rm H.c.} \} ) , \label{eq:adiabaticelimME}
\end{eqnarray}
where $\bar{\rm p}$ denotes ``not ${\rm p}$'' (i.e., ${\rm m}$ if ${\rm p=c}$ and ${\rm c}$ if ${\rm p=m}$), $\hat{H}^{\rm mc}_{\rm I}$ is given by Eq.~(\ref{eq:HmclRWA}), $\mathcal{L}^{\rm d}_{\rm mc}$ is given by Eq.~(\ref{eq:oscillatordissipation}), and the oscillator cooling/heating rates and frequency shifts due to the qubit coupling are 
\begin{subequations}
\begin{eqnarray}
\gamma^\mp_{\rm pe} & = & 2 \bar{g}^2_{\rm pq} \mathcal{R} \left[ G(\pm \omega_{\rm p} ) \right] ,  \label{eq:dampingshift} \\
\delta_{\rm p} & = & \bar{g}^2_{\rm pq} \mathcal{I} \left[ G (+\omega_{\rm p} ) - G (-\omega_{\rm p}) \right] . \label{eq:frequencyshift}
\end{eqnarray}
\end{subequations}

Here $G(\omega)$ is the fluctuation spectrum of the \emph{uncoupled} qubit,
\begin{equation}
G(\omega ) = \int^{+\infty}_0 d\tau \, e^{i\omega \tau} {\rm Tr}_{\rm q} \left[ \hat{\sigma}_x e^{ \mathcal{L}_{\rm q} \tau } \hat{\sigma}_x \rho_{\rm q} \right] , \label{eq:qubitspectrum}
\end{equation}
where the action of the qubit Liouvillian appearing in Eq.~(\ref{eq:qubitspectrum}) is given by 
\begin{equation}
\mathcal{L}_{\rm q} \rho = - \frac{i}{\hbar} [ \hat{H}^{\rm q}_{\rm I} , \rho ] + \mathcal{L}^{\rm d}_{\rm q} \rho , \label{eq:qubitLiouvillian}
\end{equation}
with $\hat{H}^{\rm q}_{\rm I}$ and $\mathcal{L}^{\rm d}_{\rm q}$ given by Eqs.~(\ref{eq:justqubitbit}) and (\ref{eq:qubitdissipation}), respectively. The $\rho_{\rm q}$ appearing in Eq.~(\ref{eq:qubitspectrum}) denotes the steady-state density matrix of the uncoupled qubit (i.e., $\mathcal{L}_{\rm q} \rho_{\rm q}=0$). 

Evaluating Eq.~(\ref{eq:qubitspectrum}) and substituting the result into Eqs.~(\ref{eq:dampingshift}) and (\ref{eq:frequencyshift}) yields
\begin{subequations}
\begin{eqnarray}
\gamma^-_{\rm pe} & = & \frac{ 4 \bar{g}^2_{\rm pq} }{\gamma_{\rm t}} \left( \frac{ \gamma_\downarrow }{ \gamma_\downarrow + \gamma_\uparrow } + \frac{\gamma_\uparrow}{\gamma_\downarrow + \gamma_\uparrow} \frac{ \gamma^2_{\rm t} }{ \gamma^2_{\rm t} + 16 \omega^2_{\rm p} } \right) , \nonumber \\
& & \label{eq:gammaminusp} \\
\gamma^+_{\rm pe} & = & \frac{ 4 \bar{g}^2_{\rm pq} }{\gamma_{\rm t}} \left(  \frac{ \gamma_\uparrow }{ \gamma_\downarrow + \gamma_\uparrow } + \frac{\gamma_\downarrow}{\gamma_\downarrow + \gamma_\uparrow} \frac{ \gamma^2_{\rm t} }{ \gamma^2_{\rm t} + 16 \omega^2_{\rm p} }  \right) , \nonumber \\
& & \label{eq:gammaplusp} \\
\delta_{\rm p} & = & 2\bar{g}^2_{\rm pq} \frac{ 4 \omega_{\rm p} }{ \gamma^2_{\rm t} + 16 \omega^2_{\rm p} } . \label{eq:deltap}
\end{eqnarray}
\end{subequations}
Now the frequency shifts have negligible impact on the steady-state provided that $\gamma^-_{\rm pe} - \gamma^+_{\rm pe} \gg \delta_{\rm p}$, which is equivalent to the requirement that $\omega_{\rm p} \gg \gamma_\downarrow$ and $\gamma_\downarrow \gg \gamma_\uparrow$. These inequalities are comfortably satisfied for our anticipated parameters, and so we henceforth neglect the frequency shifts for the analytical evaluation of the steady-state. Further, for our anticipated parameters, the cross-terms in Eq.~(\ref{eq:adiabaticelimME}) are small compared with the larger oscillator relaxation and excitation rates, and we henceforth neglect those terms. Thus, we consider the master equation
\begin{eqnarray}
\dot{\rho}_{\rm s} & = & - \frac{i}{\hbar} [ \hat{H}^{\rm mc}_{\rm I} , \rho_{\rm s} ] + \mathcal{L}^{\rm d}_{\rm mc} \rho_{\rm s}  \nonumber \\
& & + \sum_{\rm p } \left( \gamma^-_{\rm pe} \mathcal{D}[\hat{a}_{\rm p}] \rho_{\rm s} + \gamma^+_{\rm pe} \mathcal{D}[\hat{a}^\dagger_{\rm p}] \rho_{\rm s} \right) . \label{eq:projectionME}
\end{eqnarray}

Given Eq.~(\ref{eq:projectionME}), the steady-state of the reduced system is readily obtained. Setting the direct electromechanical coupling to zero ($g_{\rm mc}=0$), we find that $\langle \hat{a}^\dagger_{\rm p} \hat{a}_{\rm p} \rangle = ( \gamma^+_{\rm pe} + \gamma_{\rm p} \bar{n}_{\rm p} )/ ( \gamma^-_{\rm pe} - \gamma^+_{\rm pe} + \gamma_{\rm p} )$. That is, each oscillator mode is independently cooled due to its coupling to the driven qubit, as expected \cite{wilsonrae}. Now for the quartz-superconductor quantum electromechanical system proposed, the direct qubit-mechanics coupling is weak, such that the electrical circuit mode is directly cooled while the mechanical mode is cooled \emph{sympathetically} due to its coupling to the circuit. The general result is slightly complicated, but if we take the optimal qubit driving condition (which is $\delta_{\rm d} = \omega_{\rm p}$), and make the additional assumptions $\gamma^-_{\rm ce} , \gamma_{\rm c} \gg \gamma_{\rm m} , \gamma^+_{\rm ce}$, then we find
\begin{subequations}
\begin{eqnarray}
\langle \hat{a}^\dagger_{\rm c} \hat{a}_{\rm c} \rangle & = & \bar{n}_{\rm c} \frac{\gamma_{\rm c}}{\gamma_{\rm c} + \gamma^-_{\rm ce}} , \\
\langle \hat{a}^\dagger_{\rm m} \hat{a}_{\rm m} \rangle & = &\bar{n}_{\rm c}  \frac{\gamma_{\rm c}}{\gamma_{\rm c} + \gamma^-_{\rm ce}} \frac{ 4g^2_{\rm mc} }{ 4g^2_{\rm mc} + ( \gamma_{\rm m} + \gamma^-_{\rm me} ) ( \gamma_{\rm c} + \gamma^-_{\rm ce} ) } \nonumber \\
& & + \bar{n}_{\rm m} \frac{ ( \gamma_{\rm m} + \gamma^-_{\rm me} ) (\gamma^-_{\rm ce} + \gamma_{\rm c} ) }{ ( \gamma_{\rm m} + \gamma^-_{\rm me} ) (\gamma^-_{\rm ce} + \gamma_{\rm c} ) + 4g^2_{\rm mc} } .
\end{eqnarray}
\end{subequations}
Consequently, subject to the stated assumptions, ground-state cooling of the mechanical mode requires 
\begin{subequations}
\begin{eqnarray}
\gamma^-_{\rm ce} & > & \gamma_{\rm c} , \label{eq:req1} \\
4 g^2_{\rm mc} & >  & ( \gamma_{\rm m} + \gamma^-_{\rm me} ) (\gamma^-_{\rm ce} + \gamma_{\rm c} ) . \label{eq:req2}
\end{eqnarray}
\end{subequations}
The former is simply a necessary but not sufficient requirement for ground-state cooling of the electrical circuit mode. The latter is a requirement for ground-state cooling of the mechanical mode; the direct electromechanical coupling must be made sufficiently large compared with the product of the effective damping rates of the two oscillator modes. 

The effective parameters that emerge in the adiabatic limit in our case can be evaluated, giving (for the circuit-qubit coupling effective parameters) $\gamma^-_{\rm ce}/2\pi = 399 \, {\rm kHz}$, $\gamma^+_{\rm ce}/2\pi = 440 \, {\rm Hz}$, and $\delta_{\rm c}/2\pi = 2.0 \, {\rm kHz}$. The corresponding effective parameters for the mechanical-qubit coupling are very small: $\gamma^-_{\rm me}/2\pi = 1 \, {\rm mHz}$, $\gamma^+_{\rm me}/2\pi = 1 \, {\rm \mu Hz}$, and $\delta_{\rm m}/2\pi = 5 \, {\rm \mu Hz}$, and have a negligible impact on the dynamics of the system. However, for the sake of generality, we retain both couplings to the qubit in our subsequent analysis as the same considerations could apply to a variety of hybrid quantum systems. Now, these parameters satisfy the requirements for ground-state cooling, Eqs.~(\ref{eq:req1}) and (\ref{eq:req2}), though not by so great a margin that the results obtained in the adiabatic limit should be accepted without validation via numerical calculations.

\subsection{Sideband Picture}
The Hamiltonian (\ref{eq:timedependentHam})-(\ref{eq:dcqubitcoupling}) is explicitly time-dependent. For convenient numerical analysis, we seek a time-independent description of the system that is valid outside of the adiabatic limit. We can obtain such a description by assuming that the qubit is driven on its red sideband corresponding to the oscillator resonance frequencies (i.e.,~$\delta_{\rm d} = \omega_{\rm p}$), making an additional unitary transformation $\hat{U} = \exp \left[ i \delta_{\rm d} \hat{\sigma}_z t/2 \right]$ on (\ref{eq:timedependentHam}), and making another rotating-wave approximation on the oscillator-qubit couplings. This leads to the simple sideband Hamiltonian \cite{beaudoin},
\begin{subequations}
\begin{eqnarray}
\hat{H}_{\rm SB} & = & \hat{H}^{\rm mc}_{\rm I } + \sum_{\rm p} \hat{H}^{\rm pq}_{\rm I,RWA} , \label{eq:sidebandHam} \\
\hat{H}^{\rm pq}_{\rm I,RWA} & = & \hbar \bar{g}_{\rm pq} (\hat{a}_{\rm p} \bar{\sigma}_+ + \hat{a}^\dagger_{\rm p} \bar{\sigma}_- ) ,
\end{eqnarray}
\end{subequations}
where we stress that the Pauli operators are defined in a different frame from that in Eq.~(\ref{eq:timedependentHam}), indicated by the overbar notation. Now the steady-state of the entire system can be easily obtained by the direct numerical integration of the master equation (\ref{eq:ME}) with the Hamiltonian (\ref{eq:sidebandHam}). The sideband picture retains the description of the qubit excitation and near-resonant motional sideband, while neglecting the off-resonant motional sideband. It provides a reliable approximation provided that the resolved-sideband condition (\ref{eq:resolvedsideband}) is well-satisfied, as is the case for our proposed system.

The steady-state mechanical occupation determined numerically in the sideband picture is shown in Fig.~2, and compared with analytical results obtained in the adiabatic limit. The mechanical occupation is shown as a function of the qubit relaxation rate, for a range of electrical circuit mode intrinsic damping rates. These results demonstrate that ground-state cooling is feasible with the proposed quartz-superconductor quantum electromechanical system. We see that the analytical adiabatic limit results provide a good approximation provided that the condition (\ref{eq:adiabaticlimit}) is well-satisfied. With our assumed couplings, $\gamma_{\downarrow} \gg \bar{g}_{\rm cq}$ is satisfied on the right-hand-side of the plot, but not on the left-hand-side of the plot. To exploit the analytical results in an actual experiment, we would want to place the system into the adiabatic regime.  

\begin{figure}[h]
\begin{center}
	\includegraphics[width=0.45\textwidth]{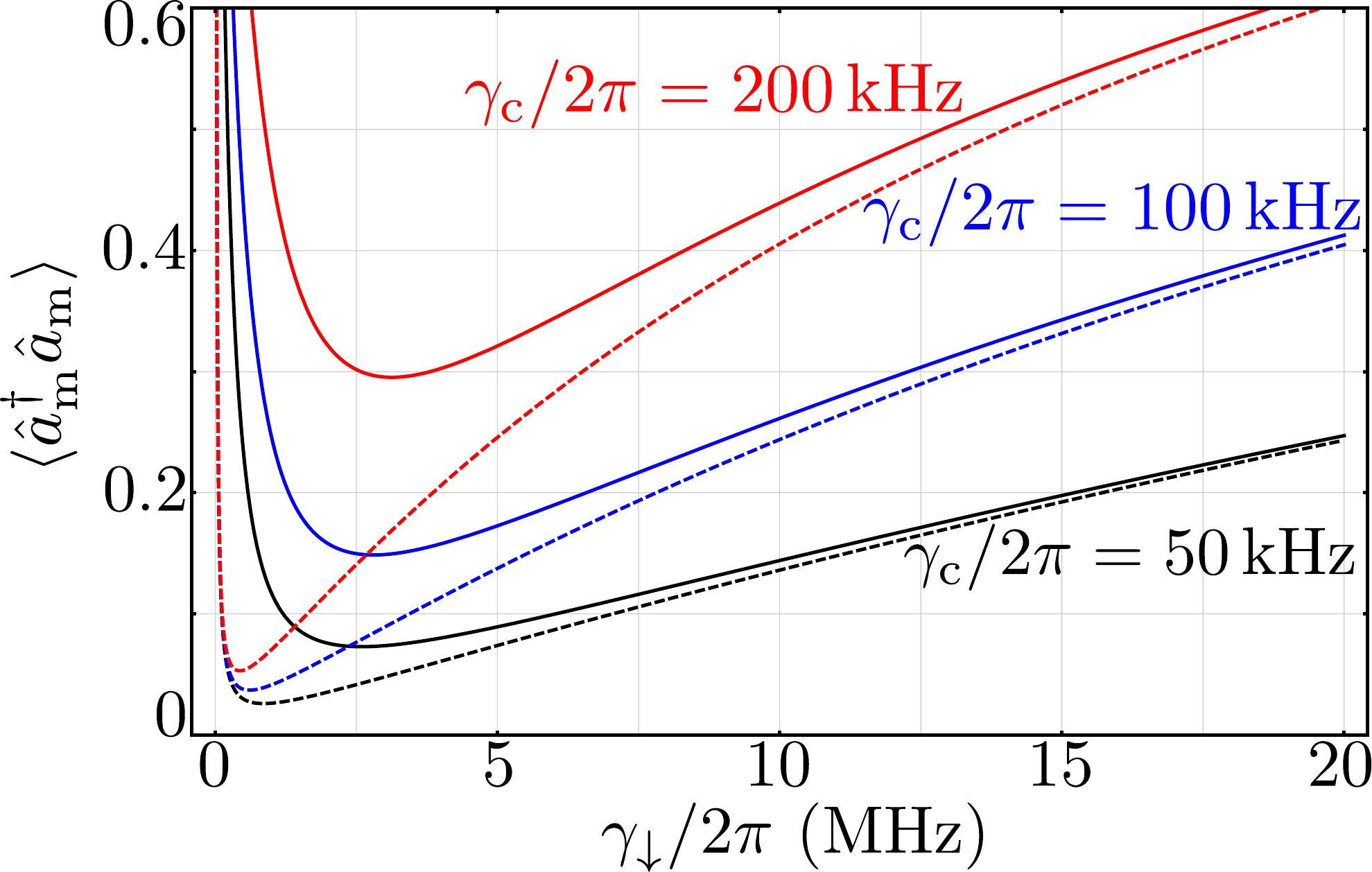} 
\end{center}
	\caption{ Steady-state mechanical occupation, $\langle \hat{a}^\dagger_{\rm m} \hat{a}_{\rm m} \rangle$, as a function of the qubit relaxation rate, $\gamma_{\downarrow}$, for $\gamma_{\rm c}/2\pi = \left\{ 50, 100, 200 \right\} \, {\rm kHz} $, corresponding to the black, blue and red curves (respectively). Numerical results obtained in the sideband picture  are shown as solid lines, and analytical results in the adiabatic limit are shown as dashed lines. Ground-state cooling is shown to be feasible with the anticipated parameters. The results from the two cases are consistent provided that $\gamma_{\downarrow} \gg \bar{g}_{\rm cq}$, corresponding to the right-hand-side of the plot. For $\gamma_{\downarrow} \sim \bar{g}_{\rm cq}$, corresponding to the left-hand-side of the plot, the adiabatic approximation breaks down and leads to an over-estimate of the cooling achievable. Increasing the $LC$ tank circuit decay rate (which is a relatively uncertain parameter) over the specified range degrades the cooling performance of the system. } 
	\label{fig:MechanicalOccupationAdiabatic}
\end{figure}

For completeness, an adiabatic elimination in this sideband picture is also given in App.~\ref{app:AdiabaticSideband}. Also note that the master equation (\ref{eq:ME}) with the Hamiltonian (\ref{eq:sidebandHam}) is closely related to the dissipative Jaynes-Cummings model, which can actually be solved analytically using continued fractions \cite{CRZ}. However, the addition of another oscillator renders this approach inapplicable here.  

\section{Qubit Spectrum}
From Sec.~\ref{sec:steadystate} it is clear that by appropriately driving the qubit, we can cool the electromechanical system. Here we show that by measuring the fluorescence of the driven qubit, we can transduce the electromechanical system. In particular, the coupling of the qubit to the oscillators results in sidebands on the qubit spectrum which are related to the steady-state number expectations of the coupled oscillators, similarly to the \emph{motional sidebands} observed in the fluorescence spectrum of a trapped ion \cite{bergquist:motionalsidebands} or a cavity optomechanical system \cite{clerk:motionalSBs}. 

This motional sideband contribution to the qubit fluorescence spectrum may be determined analytically in the limit of weak oscillator-qubit couplings using a perturbative expansion of the Liouvillian, an approach pioneered by Cirac and co-workers in the case of a trapped ion \cite{cirac:spectrumoftrappedion}. Going beyond the trapped ion case, here the qubit is coupled to \emph{two} oscillator modes, which are themselves also \emph{mutually coupled} to each other. Additionally in our case, and again in contrast to the trapped ion case, each oscillator mode is \emph{damped} into a reservoir, and that reservoir is assumed to be at some \emph{finite temperature}. The calculation that we describe here is also relevant to other experimental systems composed of one \cite{yasu} or two \cite{pirkk} thermal oscillators coupled to a qubit.  

\subsection{Adiabatic Limit}
Now the quantity that we wish to evaluate is the qubit fluorescence spectrum, given by
\begin{equation}
S[\omega ] = {\rm Re} \, \int^{+\infty}_0 dt \, e^{-i\omega t} \left\langle \hat{\sigma}_+(t) \hat{\sigma}_-(0) \right\rangle_{\rm ss} . \label{eq:basicqubitspectrum}
\end{equation}
This is to be evaluated under the Hamiltonian (\ref{eq:timedependentHam}) and the master equation (\ref{eq:ME}), such that the frequency $\omega$ in (\ref{eq:basicqubitspectrum}) is defined relative to the qubit drive frequency. 

To proceed further, we express the Hamiltonian (\ref{eq:timedependentHam}) in a Schr\"{o}dinger picture for the mechanical and electrical circuit modes, leaving
\begin{subequations}
\begin{eqnarray}
\hat{H}_{\rm SI} & = & \hat{H}^{\rm q}_{\rm I} + \hat{H}^{\rm mc}_{\rm S,RWA} + \sum_{\rm p} \hat{H}^{\rm pq}_{\rm SI} , \label{eq:labFrameHam} \\
\hat{H}^{\rm mc}_{\rm S,RWA} & = & \hbar \omega_{\rm c} \sum_{\rm p} \hat{a}^\dagger_{\rm p} \hat{a}_{\rm p} + \hbar g_{\rm mc} ( \hat{a}^\dagger_{\rm m} \hat{a}_{\rm c} + \hat{a}_{\rm m} \hat{a}^\dagger_{\rm c} ), \nonumber \\
& & \label{eq:normalmodeHam} \\
\hat{H}^{\rm pq}_{\rm SI} & = & \hbar \bar{g}_{\rm pq} ( \hat{a}_{\rm p} + \hat{a}^\dagger_{\rm p} ) \hat{\sigma}_x , \label{eq:labFrameHam3}
\end{eqnarray}
\end{subequations}
where $\hat{H}^{\rm q}_{\rm I}$ is given by Eq.~(\ref{eq:justqubitbit}).

Now we must evaluate the spectrum (\ref{eq:basicqubitspectrum}) under the Hamiltonian (\ref{eq:labFrameHam}) and the master equation (\ref{eq:ME}). To do so, we make a polaron transformation \cite{weiss} on the Pauli operators given by the unitary transformation 
\begin{subequations}
\begin{eqnarray}
\hat{U} & = & \exp \left[ i \sum_{\rm p} \tilde{g}_{\rm pq} ( \hat{a}_{\rm p} + \hat{a}^\dagger_{\rm p} ) \hat{\sigma}_z / 2 \right] , \label{eq:polarontransformation} \\
\tilde{g}_{\rm pq} & = & \bar{g}_{\rm pq}/(\gamma_\downarrow + \gamma_\uparrow ) ,
\end{eqnarray}
\end{subequations}
with $\tilde{g}_{\rm pq}$ being the sideband-reduced oscillator-qubit couplings scaled by the sum of the relaxation and excitation rates of the qubit. This transformation makes the spectrum (\ref{eq:basicqubitspectrum}) \emph{explicitly} dependent on the mechanical and electrical circuit mode operators as 
\begin{eqnarray}
S[\omega ] & = & {\rm Re} \, \int^{+\infty}_0 dt \, e^{-i\omega t} \nonumber \\
& & \times \left\langle \sigma_+ (t) e^{+ i \sum_{\rm p} \tilde{g}_{\rm pq} q_{\rm p} (t) } \sigma_- (0) e^{ - i \sum_{\rm p} \tilde{g}_{\rm pq} q_{\rm p} (0) } \right\rangle_{\rm ss} , \label{eq:transformedspectrum} \nonumber \\
\end{eqnarray}
where $q_{\rm p}(t) = a_{\rm p}(t) + a^\dagger_{\rm p}(t)$ are oscillator quadrature operators, which are invariant under the transformation (\ref{eq:polarontransformation}). Note that the \emph{post-polaron-transformation} operators are distinguished from the \emph{pre-polaron-transformation} operators by the lack of a hat; this notation shall be used throughout the remainder of the Article. The polaron-transformed Hamiltonian and master equation then take the same form as (\ref{eq:labFrameHam}) and (\ref{eq:ME}), respectively, to first-order in $\tilde{g}_{\rm pq}$, albeit in terms of the post-polaron-transformation (i.e., hatless) operators.   

Using the quantum regression theorem \cite{gardiner,swain}, the spectrum (\ref{eq:transformedspectrum}) may be expressed as
\begin{subequations}
\begin{eqnarray}
S[\omega ] & = & {\rm Re} \, \int^{+\infty}_0 dt \, e^{-i\omega t} {\rm Tr} \left[ \sigma_+ e^{i \sum_{\rm p} \tilde{g}_{\rm pq} q_{\rm p} } \mu (t) \right] , \nonumber \\
& &  \label{eq:spectrummu} \\
\mu (t) & = & e^{ \mathcal{L} t } \sigma_- e^{ -i \sum_{\rm p} \tilde{g}_{\rm pq} q_{\rm p} } \rho_{\rm ss} , \label{eq:mu}
\end{eqnarray}
\end{subequations}
where the Liouvillian $\mathcal{L}$ corresponds to the master equation (\ref{eq:ME}) with the Hamiltonian (\ref{eq:labFrameHam}) expressed in terms of post-polaron-transformation operators, and $\rho_{\rm ss}$ is the steady-state density matrix of this system (i.e., $\mathcal{L} \rho_{\rm ss} = 0$). 

In the \emph{adiabatic limit} in which the qubit is damped rapidly compared with other system rates, defined by the condition (\ref{eq:adiabaticlimit}), the $\tilde{g}_{\rm pq}$ are small parameters. Therefore, the spectrum (\ref{eq:spectrummu}) may be calculated perturbatively in $\tilde{g}_{\rm pq}$; this calculation is described in detail in App.~\ref{app:qubitspectrum}. 

Ultimately, we find that the spectrum consists of contributions from the excitation of the uncoupled qubit and the motional sideband contributions due to coupling to the oscillator modes. Thus, the qubit fluorescence spectrum may be written as 
\begin{equation}
S[\omega ] = S_{\rm q}[\omega ] + \sum_{ \rm p } S_{\rm p}[\omega ] ,  \label{eq:spectrumdecomposed} 
\end{equation}
where $S_{\rm q}[\omega ]$ is the fluorescence spectrum of the uncoupled qubit, 
\begin{eqnarray}
S_{\rm q}[\omega ] & = & {\rm Re} \, \int^{+\infty}_0 dt \, e^{-i\omega t} \langle \sigma_+(t) \sigma_- (0) \rangle_{\rm ss} \nonumber \\
& = & \frac{\gamma_\uparrow}{\gamma_\downarrow + \gamma_\uparrow} \frac{ \gamma_{\rm t}/2 }{ (\gamma_{\rm t}/2)^2 + (\omega - \delta_{\rm d})^2 } . \label{eq:usualqubit}
\end{eqnarray}
In evaluating the qubit fluctuation spectrum in Eq.~(\ref{eq:usualqubit}) we have neglected a correction first-order in $\tilde{g}_{\rm pq}$, which is assumed to be small. 

The $S_{\rm p}[\omega ]$ in Eq.~(\ref{eq:spectrumdecomposed}) are the motional sideband contributions, given by
\begin{widetext}
\begin{eqnarray}
S_{\rm p}[\omega ] & = & {\rm Re} \, \sum^{}_{\lambda'_{\rm mc}}{}^{'} \frac{ 1 }{i\omega - \lambda'_{\rm mc} } \bar{g}_{\rm pq} \sum_{\rm p'} \bar{g}_{\rm p'q} r^*( \lambda_{\rm mc}) \left[ r(\lambda_{\rm mc}) \left\langle q_{\rm p} \hat{\Pi}_{\lambda_{\rm mc}} q_{\rm p'}\right\rangle_{\rm ss} + t(\lambda_{\rm mc}) \left\langle \left[ q_{\rm p} \hat{\Pi}_{\lambda_{\rm mc}} , q_{\rm p'} \right] \right\rangle_{\rm ss} \right] ,\label{eq:oscillatorbit}
\end{eqnarray}
\end{widetext}
where ${\rm p'}$ is summed over the set $\{ {\rm m}, \, {\rm c}\}$. Here, the components of the spectrum are expressed as a sum over the \emph{eigenvalues} of the Liouvillian $\mathcal{L}$ of the master equation (\ref{eq:ME}) with the Hamiltonian (\ref{eq:labFrameHam}). More precisely, $\lambda'_{\rm mc}$ and $\lambda_{\rm mc}$ are the eigenvalues of the oscillator part of the Liouvillian, with and without (respectively) the renormalisation due to the qubit coupling, given by Eqs.~(\ref{eq:gammaminusp})-(\ref{eq:deltap}). The prime notation on the summation in Eq.~(\ref{eq:oscillatorbit}) indicates that eigenvalues with real components of the order of the qubit relaxation rate are explicitly excluded from the summation; they will not make a significant contribution since this rate is assumed to be large. The functions $r$ and $t$ appearing in Eq.~(\ref{eq:oscillatorbit}) are the uncoupled qubit correlation functions evaluated at $\lambda_{\rm mc}$, 
\begin{subequations}
\begin{eqnarray}
r(\lambda_{\rm mc} ) & = & \int^{+\infty}_0 dt \, e^{ + \lambda_{\rm mc} t } \left\langle \left[ \sigma_-(t), \sigma_x (0) \right] \right\rangle_{\rm ss } , \nonumber \\
& & \label{eq:rapp} \\
t(\lambda_{\rm mc} ) & = & \int^{+\infty}_0 dt \, e^{ - \lambda_{\rm mc} t } \left\langle \sigma_x (t) \sigma_- (0) \right\rangle_{\rm ss} \nonumber \\
& & + \int^{+\infty}_0 dt \, e^{ +\lambda_{\rm mc} t } \left\langle \sigma_x (0) \sigma_- (t) \right\rangle_{\rm ss} . \nonumber \\
& & \label{eq:tapp}
\end{eqnarray}
\end{subequations}
Eqs.~(\ref{eq:rapp}) and (\ref{eq:tapp}) are evaluated explicitly in App.~\ref{app:uncoupledqubit}. The operators $\hat{\Pi}_{\lambda_{\rm mc}}$ appearing in the moments of oscillator-space quadrature operators in Eq.~(\ref{eq:oscillatorbit}) are projection operators on the oscillator space corresponding to the oscillator-space eigenvalues $\lambda_{\rm mc}$.

Given Eqs.~(\ref{eq:spectrumdecomposed})-(\ref{eq:oscillatorbit}), our tasks are now to evaluate the moments of oscillator quadrature operators and projection operators, and then evaluate the summation in Eq.~(\ref{eq:oscillatorbit}). 

This is first performed under the assumption that
\begin{equation}
16g^2_{\rm mc} < (\gamma_{\rm c,eff} - \gamma_{\rm m,eff})^2 , \label{eq:weakcouplingassumption}
\end{equation}
where we have introduced the new \emph{effective} oscillator decay rates,
\begin{equation}
\gamma_{\rm p,eff} = \gamma_{\rm p} + \gamma^-_{\rm pe} - \gamma^+_{\rm pe} .
\end{equation}
The assumption (\ref{eq:weakcouplingassumption}), which is expected to be comfortably satisfied in our system, means that the electromechanical coupling does not affect the imaginary part of the oscillator Liouvillian eigenvalues. This assumption allows us to approximate the projection operators on the oscillator space corresponding to these eigenvalues as simply the product of the projection operators of the mechanical mode and the electrical circuit mode treated independently. Crucially, we can then assume that all the oscillator \emph{cross-correlations} (i.e., moments with ${\rm p} \neq {\rm p'}$) in (\ref{eq:oscillatorbit}) are zero. The calculation is detailed in App.~\ref{app:gmczero}. 

Setting $\delta_{\rm d} = \omega_{\rm p}$ (i.e., qubit driving on the red oscillator sideband of the qubit) and assuming $\gamma_\downarrow \gg \gamma_\uparrow$, the motional sideband contributions to the spectrum may be decomposed into \emph{upper} and \emph{lower} sideband components (relative to the qubit drive frequency). This leads to
\begin{subequations}
\begin{eqnarray}
S_{\rm p}[\omega ] & = & S^{\rm u}_{\rm p}[\omega ] + S^{\rm l}_{\rm p}[\omega ] , \label{eq:sidebandsdecomposed} \\
S^{\rm u}_{\rm p} [\omega ] & = & \frac{ 8 \bar{g}^2_{\rm pq} }{ (\gamma_{\rm t} + \tilde{\gamma}_{\rm p})^2 } \frac{ \tilde{\gamma}_{\rm p,eff} }{ 4 ( \omega - \omega_{\rm p} - \delta_{\rm p} )^2 + \tilde{\gamma}^2_{\rm p,eff} } \langle \hat{a}^\dagger_{\rm p} \hat{a}_{\rm p} \rangle_{\rm ss}  , \nonumber\\
& & \label{eq:upper} \\
S^{\rm l}_{\rm p}[\omega ] & = & \frac{ 8 \bar{g}^2_{\rm pq} }{ (\gamma_{\rm t} + \tilde{\gamma}_{\rm p} )^2 + 16 \omega^2_{\rm p} } \nonumber \\
& & \times \frac{ \tilde{\gamma}_{\rm p,eff} }{ 4 ( \omega + \omega_{\rm p} + \delta_{\rm p} )^2 + \tilde{\gamma}^2_{\rm p,eff} } \left( \langle \hat{a}^\dagger_{\rm p} \hat{a}_{\rm p} \rangle_{\rm ss} + 1 \right) , \nonumber \\
& & \label{eq:lower}
\end{eqnarray}
\end{subequations}
where the damping rates with tildes are introduced (assuming $\gamma_{\rm c} > \gamma_{\rm m}$ and $\gamma_{\rm c,eff} > \gamma_{\rm m,eff}$) via
\begin{subequations}
\begin{eqnarray}
2\tilde{\gamma}_{\rm p} & = & \gamma_{\rm m} + \gamma_{\rm c} \pm \sqrt{ ( \gamma_{\rm c} - \gamma_{\rm m} )^2 - 16 g^2_{\rm mc} } , \label{eq:decaytilderates} \\ 
2\tilde{\gamma}_{\rm p,eff} & = & \gamma_{\rm m,eff} + \gamma_{\rm c,eff} \pm \sqrt{ ( \gamma_{\rm c,eff} - \gamma_{\rm m,eff} )^2 - 16 g^2_{\rm mc} } , \nonumber \\
\label{eq:decaytilderateseff}
\end{eqnarray}
\end{subequations}
with the $+\, (-)$ signs corresponding to ${\rm p}={\rm c\, (m)}$. The motional sideband contributions for \emph{arbitrary} $\delta_{\rm d}$ are quoted in App.~\ref{app:gmczero}. 

\begin{figure}[ht]
\begin{center}
$\begin{array}{c}
	\includegraphics[width=0.45\textwidth]{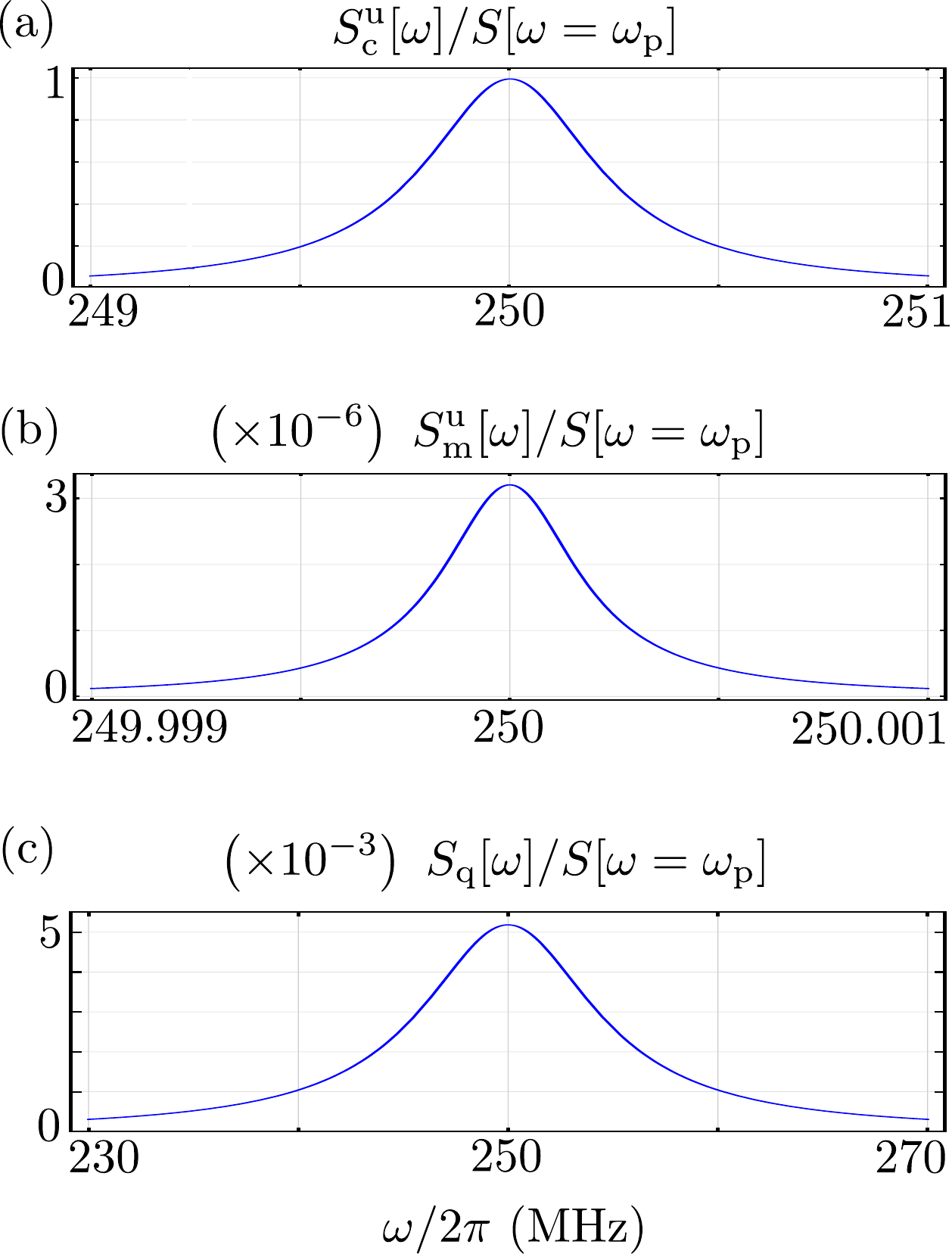} \end{array}$	
\end{center}
	\caption{ Contributions to the qubit fluorescence spectrum under the driving condition $\delta_{\rm d} = \omega_{\rm p}$, at $\omega \sim \omega_{\rm p}$ (i.e., the upper motional sideband frequency). The frequency $\omega$ is defined relative to the qubit drive frequency. The contributions are scaled by the total spectral density at the upper motional sideband $(\omega = \omega_{\rm p})$, as given by Eq.~(\ref{eq:spectrumdecomposed}). The contributions are due to: (a) coupling to the electrical circuit mode, (b) coupling to the mechanical oscillator mode, and (c) qubit excitation from its environment. Clearly, the observed spectrum is dominated by the upper motional sideband contribution from the electrical circuit. This motional sideband enables transduction of the electromechanical system. Note that the scaling of the frequency axis also varies substantially between these plots. } 
\label{fig:MotionalSBsAgain}
\end{figure}

We see that the upper motional sideband (\ref{eq:upper}) is a Lorentzian located at $\omega \approx \omega_{\rm p}$, resonant with the qubit and so is enhanced, while the lower motional sideband (\ref{eq:lower}) is a Lorentzian located at $\omega \approx -\omega_{\rm p}$, detuned by twice the oscillator resonance frequencies from the qubit resonance, and so is suppressed. Crucially, the upper motional sideband contributions are proportional to the number expectation of the oscillator modes. Assuming knowledge of the relevant system parameters, one can then measure the steady-state oscillator occupations. The contributions arising from the different oscillator modes can be distinguished by their difference in linewidths. Note that the mechanical motional sideband contribution is broadened by virtue of its direct coupling to the electrical circuit mode (and the circuit motional sideband is correspondingly narrowed), as expected from Eqs.~(\ref{eq:upper}) and (\ref{eq:decaytilderateseff}). This provides an experimental signature of the direct electromechanical coupling. 

Note that Eqs.~(\ref{eq:upper}) and (\ref{eq:lower}) also incorporate the usual motional sideband asymmetry between up-conversion and down-conversion processes which is attributable to the possibility (impossibility) of absorption (emission) processes from the quantum ground state \cite{cirac:spectrumoftrappedion}. Further, we note the similarity of each motional sideband contribution here, obtained after a number of simplifying approximations, to those obtained when the auxiliary cooling system is itself an oscillator \cite{wilsonrae2}. This similarity arises due to the fact that in the adiabatic regime ($\tilde{g}_{\rm pq} \ll 1$) the qubit decays before a scattering process is likely to attempt to excite it again, such that the distinction between a qubit and an oscillator is not very important.

Now Eq.~(\ref{eq:decaytilderateseff}) describes the hybridisation of the effective oscillator damping rates of the oscillator modes for $16g^2_{\rm mc} < (\gamma_{\rm c,eff} - \gamma_{\rm m,eff})^2$; they become equal at $16g^2_{\rm mc} = (\gamma_{\rm c,eff} - \gamma_{\rm m,eff})^2$. For $16g^2_{\rm mc} > (\gamma_{\rm c,eff} - \gamma_{\rm m,eff})^2$, the damping rates of the two oscillator modes are the same, and we see the emergence of two non-degenerate electromechanical normal modes. The evaluation of Eq.~(\ref{eq:oscillatorbit}) in the case that the condition (\ref{eq:weakcouplingassumption}) does not hold is outlined in App.~\ref{app:MotionalSideband2}.  

The uncoupled qubit and upper motional sideband contributions (electrical circuit and mechanical) to the total qubit fluorescence spectrum, as given by Eqs.~(\ref{eq:usualqubit}) and (\ref{eq:upper}) respectively, are plotted in Fig.~\ref{fig:MotionalSBsAgain}. The results are plotted for the anticipated experimental parameters listed in Sec.~\ref{sec:anticipatedparameters}. Clearly, the total qubit fluorescence spectrum is dominated by the motional sideband contribution from the coupling to the electrical circuit mode. Note that under the assumed (red sideband) driving conditions, the lower motional sidebands are suppressed by four orders of magnitude compared with the upper motional sidebands. 

The upper motional sideband should allow the direct measurement of the steady-state electrical circuit occupation, which in the parameter regime specified by Eq.~(\ref{eq:req2}), is essentially equal to the steady-state mechanical occupation. As noted above, the oscillator contributions to the spectrum are distinguishable via their linewidths, such that independent transduction of the oscillator modes in this way is possible, in principle. Unfortunately, the difference in power spectral densities would make this very challenging for the system considered here. However, the circuit motional sideband itself arises, in part, from coupling to the phononic excitations of the quartz oscillator. The interpretation of these results in terms of the $LC$ circuit and transmon as a composite transducer \cite{cernotik} for the quartz mechanical oscillator system shall be described elsewhere.

\subsection{Sideband Picture}
As for the calculation of the steady-state, an alternative to the adiabatic limit is provided by the sideband picture. The calculations in this picture are unconstrained by the adiabatic condition (\ref{eq:adiabaticlimit}), but more strongly constrained by the sideband resolution condition (\ref{eq:resolvedsideband}). It is expected that the system under consideration shall be well-described by the sideband picture. Hence, numerical calculations in this description enable us to check and assess the validity of the useful analytical results obtained in the adiabatic limit. 

The spectrum may be conveniently calculated numerically using the time-independent Hamiltonian (\ref{eq:sidebandHam}). It is given by
\begin{eqnarray}
\bar{S}[\omega ] & = & {\rm Re} \, \int^{+\infty}_0 dt \, e^{-i\omega t} \left\langle \bar{\sigma}_+(t) \bar{\sigma}_-(0) \right\rangle_{\rm ss} \nonumber \\
& = & S[\omega + \delta_{\rm d}] , \label{eq:spectruminsidebandpic}
\end{eqnarray}
where $S[\omega ]$ is as defined in Eq.~(\ref{eq:basicqubitspectrum}). The correlation function in (\ref{eq:spectruminsidebandpic}) is obtained using the quantum regression theorem \cite{swain,gardiner}, as
\begin{eqnarray}
\langle \bar{\sigma}_+ (t ) \bar{\sigma}_-(0) \rangle_{\rm ss} & = & 
\lim_{t' \rightarrow \infty} {\rm Tr} \left[ \bar{\sigma}_+ (t'+t ) \bar{\sigma}_-(t') \rho (t') \right] \nonumber \\
& = & \lim_{t' \rightarrow \infty} {\rm Tr} \left[ \bar{\sigma}_+ e^{-\mathcal{L} t} \bar{\sigma}_- \rho (t' ) e^{ \mathcal{L} t} \right] , \nonumber \\
& & 
\end{eqnarray}
where $\mathcal{L}$ denotes the Liouvillian of Eq.~(\ref{eq:ME}) with the Hamiltonian (\ref{eq:sidebandHam}). That is, the required correlation function follows from the solution of the master equation (\ref{eq:ME}) subject to the initial condition $\lim_{t' \rightarrow \infty} \bar{\sigma}_- \rho (t') $. The analytical results provide a good approximation to the numerical results in the same parameter regime that the steady-state was well-approximated by the numerical results, as indicated in Fig.~\ref{fig:MechanicalOccupationAdiabatic}.

\section{Conclusions}
Quartz BAW oscillators provide an attractive platform for the pursuit of quantum optics experiments with phonons. The difficulty of directly coupling such an oscillator to higher-frequency modes of superconducting electrical circuits led to the proposal of coupling the quartz oscillator to a superconducting transmon qubit via an intermediate $LC$ tank circuit, resonant with the quartz oscillator. Ground-state cooling of a quartz BAW oscillator mode via sideband driving of the qubit coupled to the $LC$ tank circuit is shown to be feasible. The qubit fluorescence spectrum is evaluated, with the contributions from the coupled oscillator modes determined. The mechanical and electrical circuit modes may be transduced through the observation of motional sidebands in the qubit spectrum. Cooling and measurement of high-$Q$ modes of a quartz oscillator should provide a platform for future experiments in quantum phononics. 

\section{Acknowledgments}
We wish to acknowledge support from a UNSW Canberra Early Career Researcher grant and from the Australian Research Council grant CE110001013.

\appendix

\section{Equivalent Circuit Quantisation}
\label{app:EquivalentCircuit}
We can write down the Lagrangian corresponding to the equivalent electrical circuit depicted in Fig.~1(b). The Lagrangian, $L=T-V$, consists of the kinetic and potential energy contributions,
\begin{subequations}
\begin{eqnarray}
T & = & \frac{1}{2} C_{\rm m} V^2_{\rm m} + \frac{1}{2} C_{\rm c} V^2_{\rm c} +\frac{1}{2} C_{\rm ct} V^2_{\rm ct} +  \frac{1}{2} C_{\rm t} V^2_{\rm t} , \nonumber \\
& & \\ 
V & = & \frac{ \Phi^2_{\rm m} }{2L_{\rm m}} + \frac{ \Phi^2_{\rm c} }{ 2L_{\rm c} } + E_{\rm J} \left( 1 - \cos \phi_{\rm t} \right) ,
\end{eqnarray}
\end{subequations}
respectively, where the {\rm m}, {\rm c}, and {\rm t} subscripts denote the mechanical, electrical circuit, and transmon modes, respectively, and $\phi_{\rm t} = e \Phi_{\rm t}/\hbar$ is the phase difference across the superconducting islands of the transmon. Applying Kirchhoff's voltage law around the loops of the equivalent circuit yields $V_{\rm m} = \dot{\Phi}_{\rm c} - \dot{\Phi}_{\rm m}$, $V_{\rm c} = \dot{\Phi}_{\rm c}$, $V_{\rm t} = \dot{\Phi}_{\rm t}$, and $V_{\rm ct} = \dot{\Phi}_{\rm c} - \dot{\Phi}_{\rm t}$. Given the Lagrangian in terms of the generalised coordinates $(\Phi_{\rm c}, \Phi_{\rm m}, \Phi_{\rm t})$ and the corresponding velocities, we can obtain the generalised momenta $( Q_{\rm j} = \partial L/\partial \dot{\Phi}_{\rm j} )$ and then the corresponding Hamiltonian via a Legendre transformation \cite{taylor}; $H(\Phi_{\rm j}, Q_{\rm j} ) = \sum_{\rm j} \dot{\Phi}_{\rm j} Q_{\rm j} - L ( \Phi_{\rm j}, \dot{\Phi}_{\rm j} )$. We find
\begin{eqnarray}
H & = & \frac{Q^2_{\rm c}}{ 2 \tilde{C}_{\rm c} } + \frac{ \Phi^2_{\rm c} }{2L_{\rm c}} + \frac{Q^2_{\rm m}}{ 2 \tilde{C}_{\rm m} } + \frac{ \Phi^2_{\rm m} }{2L_{\rm m}} + \frac{Q^2_{\rm t}}{2\tilde{C}_{\rm t}} \nonumber \\
& & + E_{\rm J} (1 - \cos \phi_{\rm t} ) \nonumber \\
& & + \tilde{g}_{\rm mc} Q_{\rm c} Q_{\rm m} + \tilde{\beta}_{\rm t} Q_{\rm c} Q_{\rm t} + \tilde{\beta}_{\rm t} Q_{\rm m} Q_{\rm t} , \label{eq:firstclassicalHamiltonian}
\end{eqnarray}
where the renormalised capacitances are
\begin{subequations} 
\begin{eqnarray}
\tilde{C}_{\rm c} & = & C_{\Pi}/( C_{\rm ct} + C_{\rm t} ), \\
\tilde{C}_{\rm m} & = & C_{\rm m } C_{\Pi}/(C_{\Pi} + C_{\rm \Pi m}), \\
\tilde{C}_{\rm t} & = & C_{\Pi}/(C_{\rm c} + C_{\rm ct}), 
\end{eqnarray}
\end{subequations}
with the intermediate ``capacitances'' given by
\begin{subequations}
\begin{eqnarray}
C_\Pi & = & C_{\rm c} C_{\rm ct} + C_{\rm c} C_{\rm t} + C_{\rm ct} C_{\rm t} , \\ 
C_{\rm \Pi m} & = & C_{\rm m} (C_{\rm c} + C_{\rm ct}) . 
\end{eqnarray}
\end{subequations}
The coupling constants are
\begin{subequations} 
\begin{eqnarray}
\tilde{g}_{\rm mc} & = & ( C_{\rm ct} + C_{\rm t} )/C_\Pi , \\ 
\tilde{\beta}_{\rm t} & = & C_{\rm ct}/C_\Pi . 
\end{eqnarray}
\end{subequations}

We quantise the oscillators in Eq.~(\ref{eq:firstclassicalHamiltonian}) by 
\begin{equation}
Q_{\rm p} \rightarrow \hat{Q}_{\rm p} = \sqrt{ \hbar / 2\omega_{\rm p} L_{\rm p} } ( \hat{a}_{\rm p} + \hat{a}^\dagger_{\rm p} ) , 
\end{equation}
while the transmon is quantised \cite{devoret:fluctuations} via 
\begin{subequations}
\begin{eqnarray}
Q_{\rm t} & \rightarrow & \hat{Q}_{\rm t} = 2e\hat{n} , \\ 
\phi_{\rm t} & \rightarrow & \hat{\phi} . 
\end{eqnarray}
\end{subequations}
This procedure results in the Hamiltonian (\ref{eq:firstHam}) of the Main Text, with the additional parameters 
\begin{subequations}
\begin{eqnarray}
g_{\rm mc} & = & \tilde{g}_{\rm mc}/\sqrt{4 \omega_{\rm m} L_{\rm m} \omega_{\rm c} L_{\rm c} } , \\
\beta_{\rm pt} & = & \tilde{C}_{\rm p} \tilde{\beta}_{\rm t} .    
\end{eqnarray}
\end{subequations}

The transmon is frequently approximated as a qubit \cite{koch}, as follows. The relatively large Josephson energy of the transmon ($E_{\rm J} \gg E_{\rm C}$, by definition for a transmon) restricts the phase observable to small values near zero, such that we can Taylor expand the transmon part of the Hamiltonian via $4E_{\rm C} \hat{n}^2 + E_{\rm J} \hat{\phi}^2 /2 - E_{\rm J} \hat{\phi}^4 /24$. The quadratic part of this is diagonalised by the creation and annihilation operators defined through 
\begin{subequations}
\begin{eqnarray}
\hat{\phi} & = & (2E_{\rm C}/E_{\rm J})^{1/4} ( \hat{b} + \hat{b}^\dagger ) , \\
\hat{n} & = & -i (E_{\rm J} / 32 E_{\rm C})^{1/4} ( \hat{b} - \hat{b}^\dagger ) . 
\end{eqnarray}
\end{subequations}
This results in the transmon Hamiltonian 
\begin{equation}
\hat{H}_{\rm t} = \sqrt{ 8 E_{\rm C} E_{\rm J} } \hat{b}^\dagger \hat{b} - E_{\rm C} ( \hat{b} + \hat{b}^\dagger )^4/12 , 
\end{equation}
where the quartic term is often treated perturbatively. Evaluating matrix elements of the number operator using the representation $\hat{n} = \sum_{i,j} \langle i | \hat{n} | j \rangle | i \rangle \langle j |$, truncating to the lowest two energy levels, and making a $\pi/2$ rotation of the qubit basis in the $xy$-plane, leads to the Hamiltonian (\ref{eq:SchrodingerHam}) of the Main Text. 

The effective Hamiltonian parameters follow from knowledge of the underlying equivalent electrical circuit parameters. For the equivalent electrical circuit of the quartz oscillator we expect $C_{\rm m} = 10.7\, {\rm zF}$ and $\omega_{\rm m}/2\pi = 250 \, {\rm MHz}$. These parameters are based on transmission measurements on the $77^{\rm th}$ overtone mode of a quartz BAW oscillator~\cite{rayleigh}. The transmon capacitance and the coupling capacitances are chosen to be $C_{\rm t} = 56 \, {\rm fF}$ and $C_{\rm ct} = 16 \, {\rm fF}$, respectively.

\section{Adiabatic Elimination}
\subsection{Time-Dependent Coupling}
\label{app:AdiabaticTimeDependent}
The adiabatic elimination of the qubit with the time-dependent coupling (\ref{eq:dcqubitcoupling}) may be performed using a projection operator approach \cite{gardiner,cirac:ion}, and in particular, we closely follow the approach and notation of Wilson-Rae and co-workers \cite{wilsonrae2}. First we define the projection operators 
\begin{subequations}
\begin{eqnarray}
\mathcal{P} \rho & = & {\rm Tr}_{\rm q} [ \rho ] \otimes \rho_{\rm q} , \\
\mathcal{Q} & = & 1 - \mathcal{P} ,
\end{eqnarray}
\end{subequations}
where $\rho_{\rm q}$ is the steady-state density matrix of the uncoupled qubit, as before. Introducing the formal parameter $\zeta$, the time-dependent Liouvillian corresponding to the master equation (\ref{eq:ME}) with Hamiltonian (\ref{eq:timedependentHam}) may be decomposed as 
\begin{equation}
\mathcal{L} ( t ) = \zeta^2 \mathcal{L}_{\rm q} + \zeta \mathcal{L}_1 (\zeta^2 t ) + \mathcal{L}^{\rm I}_{\rm mc} ,
\end{equation}
where $\mathcal{L}_{\rm q}$, given by Eq.~(\ref{eq:qubitLiouvillian}), acts on the qubit alone, $\mathcal{L}^{\rm I}_{\rm mc}$ acts on the mechanical oscillator and electrical circuit oscillator, and $\mathcal{L}_1$ describes the coupling between the oscillators and the qubit:
\begin{subequations}
\begin{eqnarray}
\mathcal{L}_1(\zeta^2 t ) \rho & = & \sum_{\rm p} \left( e^{ + i \omega_{\rm p} \zeta^2 t } \mathcal{L}^{(+)}_{\rm 1p} \rho + e^{-i\omega_{\rm p} \zeta^2 t} \mathcal{L}^{(-)}_{\rm 1p} \rho \right) , \nonumber \\
& &  \\
\mathcal{L}^{(+)}_{\rm 1p} \rho & = & -i\bar{g}_{\rm pq} [ \hat{\sigma}_x \hat{a}^\dagger_{\rm p} , \rho ] , \label{eq:L1plus} \\
\mathcal{L}^{(-)}_{\rm 1p} \rho & = & -i\bar{g}_{\rm pq} [ \hat{\sigma}_x  \hat{a}_{\rm p} , \rho ] , \label{eq:L1minus} \\
\mathcal{L}^{\rm I}_{\rm mc} \rho & = & - \frac{i}{\hbar} [ \hat{H}^{\rm mc}_{\rm I}, \rho ] + \mathcal{L}^{\rm d}_{\rm mc} \rho ,
\end{eqnarray}
\end{subequations}
where $\hat{H}^{\rm mc}_{\rm I}$ and $\mathcal{L}^{\rm d}_{\rm mc}$ are given by Eqs.~(\ref{eq:HmclRWA}) and (\ref{eq:oscillatordissipation}), respectively. 

Since $\mathcal{P} \rho$ is a stationary state of $\mathcal{L}_{\rm q}$, then $\mathcal{L}_{\rm q} \mathcal{P} = \mathcal{P} \mathcal{L}_{\rm q} = 0$, and subsequently $\mathcal{Q} \mathcal{L}_{\rm q} \mathcal{Q} = \mathcal{L}_{\rm q}$, and $\mathcal{P} \mathcal{L}_{\rm q} \mathcal{P} = \mathcal{Q} \mathcal{L}_{\rm q} \mathcal{P} = \mathcal{P} \mathcal{L}_{\rm q} \mathcal{Q} = 0$. The formal parameter $\zeta$ demarcates the relevant time-scales in the system, with the limit $\zeta \rightarrow \infty$ corresponding to an expansion in the ratio of fast and slow time-scales in the system. Given that we are interested in the steady-state behaviour of the system, the initial condition is irrelevant and, for simplicity, is chosen such that $\mathcal{Q} \rho (0)=0$. Integrating the differential equation for $\mathcal{Q} \rho$, we obtain a closed equation for $\mathcal{P} \rho$, given by
\begin{eqnarray}
\mathcal{P} \dot{\rho} & = & \mathcal{P} \mathcal{L}(t) \mathcal{P} + \mathcal{P} \mathcal{L}(t) \int^t_0 d\tau \, \mathcal{T}_+ \left[ e^{ \int^t_0 d\tau' \mathcal{Q} \mathcal{L}(\tau ') \mathcal{Q} } \right] \nonumber \\
& & \times \mathcal{T}_- \left[ e^{ -\int^t_0 d\tau'' \mathcal{Q} \mathcal{L}(\tau'') \mathcal{Q} } \right] \mathcal{L}(\tau ) \mathcal{P} \rho (\tau ) ,
\end{eqnarray}
where $\mathcal{T}_+$ ($\mathcal{T}_-$) is the time-ordering (anti-time-ordering) operator. In the limit $\zeta \rightarrow \infty$, we have
\begin{eqnarray}
& & \mathcal{T}_+ \left[ e^{ \int^t_0 d\tau' \mathcal{Q} \mathcal{L}(\tau ') \mathcal{Q} } \right] \mathcal{T}_- \left[ e^{ -\int^t_0 d\tau'' \mathcal{Q} \mathcal{L}(\tau'') \mathcal{Q} } \right] \nonumber \\
& = & e^{\zeta^2 \mathcal{Q} \mathcal{L}_{\rm q} \mathcal{Q}(t-\tau )} \left[ 1 + \mathcal{O}(1/\zeta ) \right] .
\end{eqnarray}
Using the operator identities listed above, as well as $\mathcal{Q}^2=\mathcal{Q}$, $\mathcal{P}^2 = \mathcal{P}$ and $\mathcal{P} \mathcal{L}^{\rm I}_{\rm mc} = \mathcal{L}^{\rm I}_{\rm mc} \mathcal{P}$, along with the change of variables $\tau'=\zeta^2 (t-\tau )$, yields 
\begin{widetext}
\begin{eqnarray}
\mathcal{P} \dot{\rho} & = & \mathcal{P} [ \zeta \mathcal{L}_1 (\zeta^2 t ) + \mathcal{L}^{\rm I}_{\rm mc} ] \mathcal{P} \rho + \mathcal{P} \mathcal{L}_1 (\zeta^2 t ) \mathcal{Q} \int^{\zeta^2 t}_0 d\tau' e^{ \mathcal{L}_{\rm q} \tau' } \mathcal{Q} \mathcal{L}_1 (\zeta^2 t - \tau') \mathcal{P} \rho (t - \tau'/\zeta^2 ) + \ldots \nonumber \\
& = & \mathcal{P} \mathcal{L}^{\rm I}_{\rm mc} \mathcal{P} \rho + \mathcal{P} \sum_{\rm p} \left( \mathcal{L}^{(+)}_{\rm 1p} \mathcal{Q} \int^{+\infty}_0 d\tau' e^{ (i\omega_{\rm p} + \mathcal{L}_{\rm q})\tau' } \mathcal{Q} \left[ \mathcal{L}^{(-)}_{\rm 1p} + \mathcal{L}^{(-)}_{1{\rm \bar{p}}} \right] \mathcal{P} \rho \right. \nonumber \\
& & \left. + \mathcal{L}^{(-)}_{\rm 1p} \mathcal{Q} \int^{+\infty}_0 d\tau' e^{ (-i\omega_{\rm p} + \mathcal{L}_{\rm q})\tau' } \mathcal{Q} \left[ \mathcal{L}^{(+)}_{\rm 1p} + \mathcal{L}^{(+)}_{1{\rm \bar{p}}} \right] \mathcal{P} \rho \right) , \label{eq:NZME}
\end{eqnarray}
\end{widetext}
where $\bar{\rm p}$ denotes ``not ${\rm p}$'' (i.e., ${\rm m}$ if ${\rm p=c}$ and ${\rm c}$ if ${\rm p=m}$), we have assumed $\omega_{\rm p} = \omega_{\rm \bar{p}}$, and in the second line we have substituted the time-dependent coupling and neglected high-frequency contributions. Using Eqs.~(\ref{eq:L1plus}) and (\ref{eq:L1minus}) leads to
\begin{subequations}
\begin{eqnarray}
& & {\rm Tr}_{\rm q} \, \left[ \mathcal{P} \mathcal{L}^{(+)}_{\rm 1p} \mathcal{Q} \int^{+\infty}_0 d\tau e^{ (i\omega_{\rm p} + \mathcal{L}_{\rm q} )\tau } \mathcal{Q} \mathcal{L}^{(-)}_{\rm 1p} \mathcal{P} \rho \right] \nonumber \\
& = & -\bar{g}^2_{\rm pq} G(+\omega_{\rm p}) \left[ \hat{a}^\dagger_{\rm p}, \hat{a}_{\rm p} \rho_{\rm s} \right] + \bar{g}^2_{\rm pq} G^*(+\omega_{\rm p}) \left[ \hat{a}_{\rm p}, \rho_{\rm s} \hat{a}^\dagger_{\rm p} \right] , \nonumber \\
& & \\
& & {\rm Tr}_{\rm q} \, \left[ \mathcal{P} \mathcal{L}^{(-)}_{\rm 1p} \mathcal{Q} \int^{+\infty}_0 d\tau e^{ (-i\omega_{\rm p} + \mathcal{L}_{\rm q} )\tau } \mathcal{Q} \mathcal{L}^{(+)}_{\rm 1p} \mathcal{P} \rho \right] \nonumber \\
& = & -\bar{g}^2_{\rm pq} G(-\omega_{\rm p}) \left[ \hat{a}_{\rm p}, \hat{a}^\dagger_{\rm p} \rho_{\rm s} \right] + \bar{g}^2_{\rm pq} G^*(-\omega_{\rm p}) \left[ \hat{a}^\dagger_{\rm p}, \rho_{\rm s} \hat{a}_{\rm p} \right] , \nonumber \\
& & \\
& & {\rm Tr}_{\rm q} \, \left[ \mathcal{P} \mathcal{L}^{(+)}_{\rm 1p} \mathcal{Q} \int^{+\infty}_0 d\tau e^{ (+i\omega_{\rm p} + \mathcal{L}_{\rm q} )\tau } \mathcal{Q} \mathcal{L}^{(-)}_{\rm 1\bar{p}} \mathcal{P} \rho \right] \nonumber \\
& = & -\bar{g}_{\rm pq} \bar{g}_{\rm \bar{p}q} \left[ \hat{a}^\dagger_{\rm p}, \hat{a}_{\rm \bar{p}} \rho_{\rm s} \right] G(+\omega_{\rm p}) \nonumber \\
& & + \bar{g}_{\rm pq} \bar{g}_{\rm \bar{p}q} [ \hat{a}_{\rm p}, \rho_{\rm s} \hat{a}^\dagger_{\rm \bar{p}} ] G^*(+\omega_{\rm p}) , \\
& & {\rm Tr}_{\rm q} \, \left[ \mathcal{P} \mathcal{L}^{(-)}_{\rm 1p} \mathcal{Q} \int^{+\infty}_0 d\tau e^{ (-i\omega_{\rm p} + \mathcal{L}_{\rm q} )\tau } \mathcal{Q} \mathcal{L}^{(+)}_{\rm 1\bar{p}} \mathcal{P} \rho \right] \nonumber \\
& = & -\bar{g}_{\rm pq} \bar{g}_{\rm \bar{p}q} [ \hat{a}_{\rm p}, \hat{a}^\dagger_{\rm \bar{p}} \rho_{\rm s} ] G(-\omega_{\rm p}) \nonumber \\
& & + \bar{g}_{\rm pq} \bar{g}_{\rm \bar{p}q} \left[ \hat{a}^\dagger_{\rm p}, \rho_{\rm s} \hat{a}_{\rm \bar{p}} \right] G^*(-\omega_{\rm p}) ,  
\end{eqnarray}
\end{subequations}
where $\rho_{\rm s} = {\rm Tr}_{\rm q} [ \mathcal{P} \rho ] = {\rm Tr}_{\rm q}[ \rho ]$, as noted in the Main Text. The equation of motion is found to be
\begin{eqnarray}
\dot{\rho}_{\rm s} & = & \mathcal{L}^{\rm I}_{\rm mc} \rho_{\rm s} - \sum_{\rm p} \bar{g}^2_{\rm pq} \{ G (+ \omega_{\rm p} ) [ \hat{a}^\dagger_{\rm p} , \hat{a}_{\rm p} \rho_{\rm s} ] \nonumber \\
& & + G(-\omega_{\rm p}) [ \hat{a}_{\rm p} , \hat{a}^\dagger_{\rm p} \rho_{\rm s} ] + {\rm H.c.} \} \nonumber \\
& & -  \sum_{\rm p} \bar{g}_{\rm pq} \bar{g}_{\rm \bar{p}q} \{ G(+\omega_{\rm p}) [ \hat{a}^\dagger_{\rm p}, \hat{a}_{\rm \bar{p}} \rho_{\rm s} ]  \nonumber \\
& & + G(-\omega_{\rm p}) [ \hat{a}_{\rm p}, \hat{a}^\dagger_{\rm \bar{p}} \rho_{\rm s} ] + {\rm H.c.} \} , \label{eq:reducedME}
\end{eqnarray}
where the uncoupled qubit fluctuation spectrum, $G(\omega )$, is given by Eq.~(\ref{eq:qubitspectrum}).

Now the qubit fluctuation spectrum under the Liouvillion $\mathcal{L}_{\rm q}$ of Eq.~(\ref{eq:qubitLiouvillian}) must be determined. To do so, we transform to a new interaction picture via $\hat{U} = \exp [i \delta_{\rm d} \hat{\sigma}_z t /2 ]$ such that the Pauli operators in the new picture are $\hat{\sigma}^{\rm n}_\pm = \hat{\sigma}_\pm e^{\mp i \delta_{\rm d} t}$. The required qubit fluctuation spectrum, Eq.~(\ref{eq:qubitspectrum}) becomes (if we neglect rapidly oscillating terms), 
\begin{eqnarray}
G(\omega ) & = & \int^{+\infty}_0 d\tau \, e^{i\omega \tau} \left( {\rm Tr}_{\rm q} [ \hat{\sigma}^{\rm n}_+ e^{ \mathcal{L}^{\rm d}_{\rm q} \tau } \hat{\sigma}^{\rm n}_- \rho^{\rm n}_{\rm q} ] e^{ + i \delta_{\rm d} \tau } \right. \nonumber \\
& & \left. + {\rm Tr}_{\rm q} [ \hat{\sigma}^{\rm n}_- e^{ \mathcal{L}^{\rm d}_{\rm q} \tau } \hat{\sigma}^{\rm n}_+ \rho^{\rm n}_{\rm q} ] e^{ - i \delta_{\rm d} \tau } \right) , \label{eq:Gomega}
\end{eqnarray}
where the ${\rm n}$ superscript denotes operators in the newly defined interaction picture. One can write out the Maxwell-Bloch equations corresponding to $\mathcal{L}^{\rm d}_{\rm q}$ of Eq.~(\ref{eq:qubitdissipation}), solve and apply the quantum regression theorem \cite{scully,woolley} to find the required steady-state correlation functions, $\langle \hat{\sigma}^{\rm n}_+(\tau) \hat{\sigma}^{\rm n}_- (0 ) \rangle_{\rm ss} = ( \gamma_\uparrow /( \gamma_\downarrow + \gamma_\uparrow )) e^{ - \gamma_{\rm t} \tau /2 }$ and $\langle \hat{\sigma}^{\rm n}_-(\tau) \hat{\sigma}^{\rm n}_+ (0 ) \rangle_{\rm ss} = ( \gamma_\downarrow /( \gamma_\downarrow + \gamma_\uparrow )) e^{ - \gamma_{\rm t} \tau /2 }$. Substituting these correlations into Eq.~(\ref{eq:Gomega}) yields
\begin{eqnarray}
G(\omega ) & = & \frac{1}{\gamma_\downarrow + \gamma_\uparrow} \left[ \frac{\gamma_\uparrow}{ \gamma_{\rm t} /2 - i (\omega + \delta_{\rm d}) } \right. \nonumber \\
& & \left. + \frac{ \gamma_\downarrow }{ \gamma_{\rm t}/2 - i (\omega - \delta_{\rm d}) } \right] . \label{eq:finalqubitfluctuations} 
\end{eqnarray}
Substituting Eq.~(\ref{eq:finalqubitfluctuations}) into Eq.~(\ref{eq:reducedME}) leads to Eq.~(\ref{eq:adiabaticelimME}) of the Main Text. 

\subsection{Sideband Picture}
\label{app:AdiabaticSideband}

For completeness, the adiabatic elimination of the qubit is also performed in the sideband picture. This is achieved by expanding the qubit part of the density matrix and eliminating coherences \cite{warszawski}. Starting at the master equation (\ref{eq:ME}) with the Hamiltonian (\ref{eq:sidebandHam}), we explicitly expand the system density matrix over the qubit Hilbert space as $\rho = \rho_{00} \otimes |0\rangle \langle 0 | + \rho_{01} \otimes |0 \rangle \langle 1 | + \rho^\dagger_{01} \otimes | 1\rangle \langle 0 | + \rho_{11} \otimes | 1 \rangle \langle 1 | $, where the operators $\rho_{mn}$ are defined on the Hilbert space of the two oscillators. Substituting this ansatz into Eq.~(\ref{eq:ME}) and equating operators on the qubit space yields the system of equations,
\begin{subequations}
\begin{eqnarray}
\dot{\rho}_{00} & = & \mathcal{L}^{\rm I,R}_{\rm mc} \rho_{00} - i\sum_{\rm p} \bar{g}_{\rm pq} ( \hat{a}^\dagger_{\rm p} \rho^\dagger_{01} - \rho_{01} \hat{a}_{\rm p} ) + \gamma_\downarrow \rho_{11} \nonumber \\
& & - \gamma_\uparrow \rho_{00} , \\
\dot{\rho}_{01} & = & \mathcal{L}^{\rm I,R}_{\rm mc} \rho_{01} - i \sum_{\rm p} \bar{g}_{\rm pq} (  \hat{a}^\dagger_{\rm p} \rho_{11} - \rho_{00} \hat{a}^\dagger_{\rm p} ) - \frac{\gamma_{\rm t}}{2} \rho_{01} , \nonumber \\
& &  \label{eq:rho01} \\
\dot{\rho}^\dagger_{01} & = & \mathcal{L}^{\rm I,R}_{\rm mc} \rho^\dagger_{01} - i  \sum_{\rm p} \bar{g}_{\rm pq} ( \hat{a}_{\rm p} \rho_{00} - \rho_{11} \hat{a}_{\rm p} ) - \frac{\gamma_{\rm t}}{2} \rho^\dagger_{01} , \nonumber \\
& & \\
\dot{\rho}_{11} & = & \mathcal{L}^{\rm I,R}_{\rm mc} \rho_{11} - i \sum_{\rm p} \bar{g}_{\rm pq} ( \hat{a}_{\rm p} \rho_{01} - \rho^\dagger_{01} \hat{a}^\dagger_{\rm p} ) - \gamma_\downarrow \rho_{11} \nonumber \\
& & + \gamma_\uparrow \rho_{00} ,
\end{eqnarray}
\end{subequations}
where 
\begin{equation}
\mathcal{L}^{\rm I,R}_{\rm mc} \rho = -(i/\hbar)[ \hat{H}^{\rm mc}_{\rm I} , \rho ] + \mathcal{L}^{\rm d}_{\rm mc} \rho , 
\end{equation}
see Eqs.~(\ref{eq:HmclRWA}) and (\ref{eq:oscillatordissipation}). The equation of motion for the reduced density matrix of the two oscillators is obtained by tracing the master equation over the qubit, 
\begin{equation}
\dot{\rho}_s = {\rm Tr_{\rm q}} \, [ \dot{\rho} ] = \dot{\rho}_{00} + \dot{\rho}_{11}, 
\end{equation}
for which we find
\begin{equation}
\dot{\rho}_s = \mathcal{L}^{\rm I,R}_{\rm mc} \rho_s - i \sum_{\rm p} \bar{g}_{\rm pq} ( [ \hat{a}^\dagger_{\rm p} , \rho^\dagger_{01} ] + [ \hat{a}_{\rm p} , \rho_{01} ] )  . \label{eq:tracedmasterequation}
\end{equation}
We can eliminate $\rho_{01}$ from (\ref{eq:tracedmasterequation}) by first setting Eq.~(\ref{eq:rho01}) to zero and solving for $\rho_{01}$. This yields 
\begin{equation}
\left( \mathcal{L}^{\rm I,R}_{\rm mc} - \gamma_{\rm t}/2 \right) \rho_{01} = i \sum_{\rm p} \bar{g}_{\rm pq} (\hat{a}^\dagger_{\rm p} \rho_{11} - \rho_{00} \hat{a}_{\rm p}). 
\end{equation}
Since the qubit is rapidly damped, we can make the approximations $\rho_{00} \sim \rho_s$ and $\rho_{11} \sim 0$. Therefore, $\rho_{01} = -i \sum_{\rm p} \bar{g}_{\rm pq} \mathcal{M} \rho_s \hat{a}^\dagger_{\rm p}$ where $\mathcal{M} = \left( \mathcal{L}^{\rm I,R}_{\rm mc} - \gamma_{\rm t}/2 \right)^{-1}$. The resulting master equation over the space of two oscillators is
\begin{eqnarray}
\dot{\rho}_s & = & \mathcal{L}^{\rm I,R}_{\rm mc} \rho_s + \sum_{\rm p,p'} \bar{g}_{\rm pq} \bar{g}_{\rm p'q} ( [ \hat{a}_{\rm p} , \mathcal{M} \rho_s \hat{a}^\dagger_{p'} ]  \nonumber \\
& & - [ \hat{a}^\dagger_{\rm p} , \hat{a}_{\rm p'} \rho_s \mathcal{M} ] ) . \label{eq:MEM}
\end{eqnarray}
We can write out the Neumann series expansion of the operator $\mathcal{M}$, giving $\mathcal{M} = - \sum^{+\infty}_{k=0} ( 2/\gamma_{\rm t} )^{k+1} ( \mathcal{L}^{\rm I,R}_{\rm mc})^k$. Substituting this into Eq.~(\ref{eq:MEM}) and truncating the Neumann series at zeroth-order yields the linear, Markovian master equation (\ref{eq:MEadiabatic}). 

It results in the master equation for the reduced density matrix describing the dynamics of the mechanical and electrical circuit oscillators,  
\begin{eqnarray}
\dot{\rho}_{\rm s} & = & - \frac{i}{\hbar} [ \hat{H}^{\rm mc}_{\rm I}, \rho ] + \mathcal{L}^{\rm d}_{\rm mc} \rho_{\rm s} + \sum_{\rm p} \left[ \frac{ 4\bar{g}^2_{\rm pq} }{\gamma_{\rm t}} \mathcal{D} [ \hat{a}_{\rm p} ] \rho_{\rm s} \right. \nonumber \\
& & \left. + \frac{2 \bar{g}_{\rm pq} \bar{g}_{\rm \bar{p} q} }{ \gamma_{\rm t} } \left( [ \hat{a}_{\rm p} , \rho_{\rm s} \hat{a}^\dagger_{ \bar{p} } ]  + {\rm H.c.} \right) \right] . \label{eq:MEadiabatic}
\end{eqnarray} 
The steady-state of Eq.~(\ref{eq:MEadiabatic}) is readily obtained, and compared with the steady-state of Eq.~(\ref{eq:adiabaticelimME}).

\section{Qubit Spectrum}
\label{app:qubitspectrum}

\subsection{Perturbative Approach}
\label{app:MotionalSideband1}

Expanding the spectrum of Eqs.~(\ref{eq:spectrummu})-(\ref{eq:mu}) to first-order in the (small) parameters $\tilde{g}_{\rm pq}$ we get
\begin{eqnarray}
S[\omega ] & = & {\rm Re} \, ( {\rm Tr}_{\rm q} \left[ \sigma_+ {\rm Tr}_{\rm mc} \left[ \tilde{\mu} (\omega ) \right] \right]  \nonumber \\
& & + i \sum_{\rm p} \tilde{g}_{\rm pq} {\rm Tr}_{\rm q} \left[ \sigma_+ {\rm Tr}_{\rm mc} \left[ q_{\rm p} \tilde{\mu} (\omega ) \right] \right] ) , \label{eq:spectrumLDL}
\end{eqnarray}
where $\tilde{\mu} (\omega ) = \int^{+\infty}_0 dt \, e^{-i\omega t} \mu (t)$, ${\rm Tr}_{\rm mc}[\ldots]$ denotes a partial trace over the mechanical and circuit oscillator spaces, and ${\rm Tr}_{\rm q}[\ldots ]$ denotes a partial trace over the qubit space. Now $\mu (t)$, given by Eq.~(\ref{eq:mu}), satisfies the master equation (\ref{eq:ME}). Tracing over the oscillator modes it follows that
\begin{eqnarray}
\frac{d}{dt} {\rm Tr}_{\rm mc} [ \mu (t) ] & = & \mathcal{L}_{\rm q} {\rm Tr}_{\rm mc} [ \mu (t) ] \nonumber \\
& & - i \sum_{\rm p} \bar{g}_{\rm pq} [ \sigma_x, {\rm Tr}_{\rm mc} [ q_{\rm p} \mu ] ] . \label{eq:ddttrace}
\end{eqnarray}
Taking the Laplace transform of Eq.~(\ref{eq:ddttrace}) and subsequently setting $s=i\omega$ yields 
\begin{eqnarray}
{\rm Tr}_{\rm mc} [ \tilde{\mu}(\omega ) ] & = & (i\omega - \mathcal{L}_{\rm q})^{-1} ( {\rm Tr}_{\rm mc}[\mu (0)]  \nonumber \\
& & - i \sum_{\rm p} \bar{g}_{\rm pq} [ \sigma_x , {\rm Tr}_{\rm mc} [ q_{\rm p} \tilde{\mu}(\omega ) ] ] ) . \nonumber \\
& & \label{eq:tracemu}
\end{eqnarray}
Substituting Eq.~(\ref{eq:tracemu}) into Eq.~(\ref{eq:spectrumLDL}), the spectrum may be written in the form
\begin{eqnarray}
S[\omega ] & = & {\rm Re} \, \left( {\rm Tr}_{\rm q} [\sigma_+ (i\omega - \mathcal{L}_{\rm q})^{-1} {\rm Tr}_{\rm mc} [ \mu (0) ] \right) \nonumber \\
& & + \sum_{\rm p} {\rm Re} \, \left[ i \left( \tilde{g}_{\rm pq} {\rm Tr}_{\rm q} [ \sigma_+ {\rm Tr}_{\rm mc} [ q_{\rm p} \tilde{\mu} (\omega ) ] ] \right. \right. \nonumber \\
& & \left. \left. - \bar{g}_{\rm pq} {\rm Tr}_{\rm q} [ \sigma_+ (i\omega - \mathcal{L}_{\rm q})^{-1} [ \sigma_x , {\rm Tr}_{\rm mc} [q_{\rm p} \tilde{\mu} (\omega ) ] ] ] \right) \right] . \nonumber \\
& & \label{eq:spectrumexpanded}
\end{eqnarray}

The first term in Eq.~(\ref{eq:spectrumexpanded}) is, neglecting small corrections of linear and higher order in $\tilde{g}_{\rm pq}$, given by 
\begin{equation}
S_{\rm q}[\omega ] = {\rm Re}\, ( {\rm Tr}_{\rm q} [ \sigma_+ (i\omega - \mathcal{L}_{\rm q})^{-1} \sigma_- \rho^{\rm q}_{\rm ss}]) \label{eq:simplequbitspectrum}
\end{equation}
where $\rho^{\rm q}_{\rm ss} = {\rm Tr}_{\rm mc} [\rho_{\rm ss}]$ is the steady-state qubit density matrix. Neglecting a small correction in $\tilde{g}_{\rm pq}$, $\rho^{\rm q}_{\rm ss}$ is equivalent to $\rho_{\rm q}$, the steady-state density matrix of the uncoupled qubit. Then Eq.~(\ref{eq:simplequbitspectrum}) is simply the fluorescence spectrum of the uncoupled qubit, given explicitly in Eq.~(\ref{eq:usualqubit}) of the Main Text. Since, in the adiabatic limit, the oscillator quadratures $q_{\rm p}$ are damped much more slowly than the qubit, the ${\rm Tr}_{\rm mc}[q_{\rm p} \tilde{\mu} (\omega ) ]$ terms will give rise to narrow motional sidebands at the eigenfrequencies of the oscillator part of the Hamiltonian. From Eq.~(\ref{eq:spectrumexpanded}), the motional sideband contributions to the qubit spectrum are given by
\begin{eqnarray}
S_{\rm p}[\omega ] & = & {\rm Re} \, \left( i \tilde{g}_{\rm pq} {\rm Tr}_{\rm q} [ \sigma_+ {\rm Tr}_{\rm mc}[ q_{\rm p} \tilde{\mu}(\omega ) ]] \right. \nonumber \\
& & \left. - i \bar{g}_{\rm pq} {\rm Tr}_{\rm q} [ \sigma_+ (i\omega - \mathcal{L}_{\rm q})^{-1} [ \sigma_x , {\rm Tr}_{\rm mc} [ q_{\rm p} \tilde{\mu} (\omega ) ] ]] \right) . \nonumber \\
& &  \label{eq:Spomega}
\end{eqnarray}
Eq.~(\ref{eq:Spomega}) crucially depends on the quantity ${\rm Tr}_{\rm mc} [ q_{\rm p} \tilde{\mu} (\omega ) ] $. Inserting the definition of $\mu (t)$, given by Eq.~(\ref{eq:mu}), into $\tilde{\mu} (\omega )$, evaluating the Laplace transform of the exponentiated Liouvillian and setting $s=i\omega$ gives 
\begin{eqnarray}
& & {\rm Tr}_{\rm mc} \left[ q_{\rm p} \tilde{\mu} (\omega ) \right] \nonumber \\ 
& = & {\rm Tr}_{\rm mc} \left[ q_{\rm p} (i\omega - \mathcal{L})^{-1} \sigma_- e^{ -i\sum_{\rm p'} \tilde{g}_{\rm p'q} q_{\rm p'} } \rho_{\rm ss} \right] . \nonumber \\
& & \label{eq:spectraldecomp}
\end{eqnarray}

\subsection{Spectral Decomposition of Liouvillian}
\label{app:spectraldecomposition}
The quantity in Eq.~(\ref{eq:spectraldecomp}) can be evaluated using a spectral decomposition of the Liouvillian,
\begin{eqnarray}
& & {\rm Tr}_{\rm mc} \left[ q_{\rm p} \tilde{\mu} (\omega ) \right] \nonumber \\
& = & \sum_\lambda \frac{1}{i\omega - \lambda} {\rm Tr}_{\rm mc} \left[ q_{\rm p} \hat{\Pi}_\lambda \sigma_- e^{ -i\sum_{\rm p'} \tilde{g}_{\rm p'q} q_{\rm p'} } \rho_{\rm ss} \right] , \nonumber \\
& & \label{eq:tracespectral}
\end{eqnarray}
where $\lambda$ are the eigenvalues of the Liouvillian $\mathcal{L}$ and $\hat{\Pi}_\lambda$ are projectors onto the subspace spanned by the eigenvectors corresponding to the eigenvalue $\lambda$. 

The peaks of Eq.~(\ref{eq:tracespectral}) will be at $\omega \sim {\rm Im} (\lambda )$, and the most important contributions to these peaks will be from terms in which the real parts of $\lambda$ are small. Now to zeroth-order in $\tilde{g}_{\rm pq}$ the oscillators are decoupled from the qubit, and the dynamics of the system are governed by the master equation [c.f. Eqs.~(\ref{eq:ME}), (\ref{eq:qubitLiouvillian}) and (\ref{eq:labFrameHam})],
\begin{eqnarray}
\frac{d}{dt} \rho^0 & = & -\frac{i}{\hbar } [ \hat{H}^{\rm mc}_{\rm S,RWA} , \rho^0 ] + \mathcal{L}^{\rm d}_{\rm mc} \rho^0 + \mathcal{L}_{\rm q} \rho^0 \nonumber \\
& \equiv & \mathcal{L}^{\rm S,R}_{\rm mc} \rho^0 + \mathcal{L}_{\rm q} \rho^0 \equiv \mathcal{L}_0 \rho^0 .
\end{eqnarray} 
Consequently, the eigenvalues of $\mathcal{L}_0$ are of the form $\lambda^0 = \lambda_{\rm mc} + \lambda_{\rm q}$, where $\lambda_{\rm mc}$ are the eigenvalues of the Liouvillian $\mathcal{L}^{\rm S,R}_{\rm mc}$ and $\lambda_{\rm q}$ are the eigenvalues of the Liouvillian $\mathcal{L}_{\rm q}$. The projection operator $\hat{\Pi}^0_\lambda$ onto the subspace spanned by $\lambda^0$ is given by $\hat{\Pi}^0_\lambda = \hat{\Pi}_{\lambda_{\rm mc}} \otimes \hat{\Pi}_{\lambda_{\rm q}}$, where $\hat{\Pi}_{\lambda_{\rm mc}}$ and $\hat{\Pi}_{\lambda_{\rm q}}$ are the projection operators corresponding to the eigenvalues $\lambda_{\rm mc}$ and $\lambda_{\rm q}$, respectively.

Now we only retain terms in Eq.~(\ref{eq:tracespectral}) corresponding to $\lambda_{\rm q}=0$, since other eigenvalues will have large real parts of the order of the qubit relaxation rate. Since we are interested in the sideband spectrum, we also neglect the small correction to the central peak corresponding to the eigenvalue $\lambda_{\rm mc}$ having zero imaginary part. The exclusion of these terms from the summation in Eq.~(\ref{eq:tracespectral}) shall subsequently be denoted using a prime notation.

Accounting for the oscillator-qubit couplings to first-order in $\tilde{g}_{\rm pq}$, we can write the corrected Liouvillian as $\mathcal{L}=\mathcal{L}_0 + \mathcal{L}_1$ where $\mathcal{L}_1$ is defined by its action 
\begin{equation}
\mathcal{L}_1 \rho = -\frac{i}{\hbar } \sum_{\rm p}[ \hat{H}^{\rm pq}_{\rm SI}, \rho ], \label{eq:L1}
\end{equation}
with $\hat{H}^{\rm pq}_{\rm SI}$ given by Eq.~(\ref{eq:labFrameHam3}). The corresponding corrections to the steady-state density matrix, eigenvalues and projection operators are $\rho_{\rm ss} = \rho^0_{\rm ss} + \rho^1_{\rm ss}$, $\lambda = \lambda^0 + \lambda^1$, and $\hat{\Pi}_\lambda = \hat{\Pi}^0_\lambda + \hat{\Pi}^1_\lambda$, respectively. Expanding the numerator of Eq.~(\ref{eq:tracespectral}) to first-order in $\tilde{g}_{\rm pq}$, substituting these expressions and dropping terms above first-order in $\tilde{g}_{\rm pq}$, we find
\begin{eqnarray}
{\rm Tr}_{\rm mc} \left[ q_{\rm p} \tilde{\mu} (\omega ) \right] & = & \sum^{}_{\lambda}{}^{'} \frac{1}{i\omega - \lambda} {\rm Tr}_{\rm mc} \left[ q_{\rm p} \hat{\Pi}^0_\lambda \sigma_- \rho^0_{\rm ss} \right. \nonumber \\
& & - i \sum_{\rm p'} \tilde{g}_{\rm p'q} q_{\rm p} \hat{\Pi}^0_\lambda \sigma_- q_{\rm p'} \rho^0_{\rm ss}  \nonumber \\
& & \left. + q_{\rm p} \hat{\Pi}^0_\lambda \sigma_- \rho^1_{\rm ss} + q_{\rm p} \hat{\Pi}^1_\lambda \sigma_- \rho^0_{\rm ss} \right] . \nonumber \\
& & \label{eq:tracefirstorder}
\end{eqnarray}

\subsection{Evaluation of First-order Corrections}
\label{app:firstorder}
In order to simplify Eq.~(\ref{eq:tracefirstorder}) further we must determine the first-order corrections in terms of known quantities, namely, the zeroth-order results and the first-order Liouvillian. Using standard quantum mechanical perturbation theory \cite{sakurai}, the first-order correction to the zeroth-order projection operator is
\begin{equation}
\hat{\Pi}^1_\lambda = \hat{\Pi}^0_\lambda \mathcal{L}_1 \frac{ 1 - \hat{\Pi}^0_\lambda }{ \lambda_0 - \mathcal{L}_0 } + \frac{ 1 - \hat{\Pi}^0_\lambda }{ \lambda_0 - \mathcal{L}_0 } \mathcal{L}_1 \hat{\Pi}^0_\lambda . \label{eq:Pi1}
\end{equation}
Now to first-order in $\tilde{g}_{\rm pq}$, the steady-state satisfies $\mathcal{L} \rho_{\rm ss} = \mathcal{L}_0 \rho^1_{\rm ss} + \mathcal{L}_1 \rho^0_{\rm ss}=0$. Applying the projection operator $\hat{\Pi}_{\lambda_{\rm mc}}$ to this equation, and using the properties $\hat{\Pi}_{\lambda_{\rm mc}} \mathcal{L}_0 = \mathcal{L}_0 \hat{\Pi}_{\lambda_{\rm mc}} = ( \lambda_{\rm mc} + \mathcal{L}_{\rm q} ) \hat{\Pi}_{\lambda_{\rm mc}}$, we find that
\begin{equation}
\hat{\Pi}_{\lambda_{\rm mc}} \rho^1_{\rm ss} = - (\lambda_{\rm mc} + \mathcal{L}_{\rm q})^{-1} \hat{\Pi}_{\lambda_{\rm mc}} \mathcal{L}_1 \rho^0_{\rm ss} . \label{eq:rho1}
\end{equation}
Provided $\lambda_{\rm q} \neq 0$, recall that we explicitly exclude the zero eigenvalue from our summation of Eq.~(\ref{eq:tracefirstorder}), we shall also use the property that for any operator $\hat{X}$, 
\begin{eqnarray}
\hat{\Pi}_{\lambda_{\rm q}} \hat{X} & = & \rho_{\rm q} {\rm Tr}_{\rm q} [ \hat{X} ] , \label{eq:Pilambdaq}
\end{eqnarray}
where $\rho_{\rm q} = {\rm Tr}_{\rm mc} [ \rho^0_{\rm ss} ]$ is the steady-state qubit density matrix of the uncoupled qubit (i.e., $\mathcal{L}_{\rm q} \rho_{\rm q}=0$). Using $\hat{\Pi}^0_\lambda = \hat{\Pi}_{\lambda_{\rm mc}} \otimes \hat{\Pi}_{\lambda_{\rm q}}$, and substituting Eqs.~(\ref{eq:Pi1}), (\ref{eq:rho1}) and (\ref{eq:Pilambdaq}) into Eq.~(\ref{eq:tracefirstorder}) leads to 
\begin{widetext}
\begin{eqnarray}
{\rm Tr}_{\rm mc} \left[ q_{\rm p} \tilde{\mu} (\omega ) \right] & = & -i \rho_{\rm q} \sum^{}_{\lambda}{}^{'} \frac{1 }{i\omega - \lambda} \sum_{\rm p'} \tilde{g}_{\rm p'q} {\rm Tr}\, \left[ q_{\rm p} \hat{\Pi}_{\lambda_{\rm mc}} q_{\rm p'} \sigma_- \rho^0_{\rm ss} \right] \nonumber \\
& & + \rho_{\rm q} \sum^{}_{\lambda}{}^{'} \frac{1}{i\omega - \lambda} {\rm Tr}\, \left[ q_{\rm p} \hat{\Pi}_{\lambda_{\rm mc}} \mathcal{L}_1 (\lambda_{\rm mc} - \mathcal{L}_{\rm q})^{-1} \sigma_- \rho^0_{\rm ss} - q_{\rm p} \hat{\Pi}_{\lambda_{\rm mc}} \sigma_- (\lambda_{\rm mc} + \mathcal{L}_{\rm q})^{-1} \mathcal{L}_1 \rho^0_{\rm ss} \right] . \nonumber \\
& & \label{eq:53}
\end{eqnarray}
\end{widetext}
We stress that the trace operators in Eq.~(\ref{eq:53}) are taken over the qubit-oscillator-oscillator space of the whole system. Evaluating Eq.~(\ref{eq:53}) term-by-term, writing $\rho^0_{\rm ss} = \rho_{\rm q} \otimes \rho_{\rm mc}$, and using Eq.~(\ref{eq:L1}) we obtain the contributions
\begin{widetext}
\begin{subequations}
\begin{eqnarray}
\sum_{\rm p'} \tilde{g}_{\rm p'q} {\rm Tr} [ q_{\rm p} \hat{\Pi}_{\lambda_{\rm mc}} q_{\rm p'} \bar{\sigma}_- \rho^0_{\rm ss} ] & = & \sum_{\rm p'} \tilde{g}_{\rm p'q} \langle q_{\rm p} \hat{\Pi}_{\lambda_{\rm mc}} q_{\rm p'} \rangle_{\rm ss} \langle \sigma_- \rangle_{\rm ss} , \label{eq:54a} \\ 
{\rm Tr} \, [ q_{\rm p} \hat{\Pi}_{\lambda_{\rm mc}} \mathcal{L}_1 (\lambda_{\rm mc} - \mathcal{L}_{\rm q})^{-1} \sigma_- \rho^0_{\rm ss} ] & = & -i \sum_{\rm p'} \bar{g}_{\rm p'q} \left\langle \left[ q_{\rm p} \hat{\Pi}_{\lambda_{\rm mc}} , q_{\rm p'} \right] \right\rangle_{\rm ss} {\rm Tr}_{\rm q} \left[ \sigma_x (\lambda_{\rm mc} - \mathcal{L}_{\rm q})^{-1} \sigma_- \rho_{\rm q} \right] , \nonumber \\
& & \\
{\rm Tr} \, [ -q_{\rm p} \hat{\Pi}_{\lambda_{\rm mc}} \sigma_- (\lambda_{\rm mc} + \mathcal{L}_{\rm q})^{-1} \mathcal{L}_1 \rho^0_{\rm ss} ] & = & i \sum_{\rm p'} \bar{g}_{\rm p'q} \left[ \langle q_{\rm p} \hat{\Pi}_{\lambda_{\rm mc}} q_{\rm p'} \rangle_{\rm ss} {\rm Tr}_{\rm q} \left[ \sigma_- ( \lambda_{\rm mc} + \mathcal{L}_{\rm q} )^{-1} [ \sigma_x , \rho_{\rm q} ] \right] \right. \nonumber \\    
& & \left. + \left\langle \left[ q_{\rm p} \hat{\Pi}_{\lambda_{\rm mc}} , q_{\rm p'} \right] \right\rangle_{\rm ss} {\rm Tr}_{\rm q} \left[ \sigma_- ( \lambda_{\rm mc} + \mathcal{L}_{\rm q} )^{-1} \rho_{\rm q} \sigma_x \right] \right] . \label{eq:54c}
\end{eqnarray}
\end{subequations}
\end{widetext}
Note that the expectations that arise in Eqs.~(\ref{eq:54a})-(\ref{eq:54c}) are expectations evaluated either in the oscillator space or in the qubit space, but not across the Hilbert space of the entire system. Substituting Eqs.~(\ref{eq:54a})-(\ref{eq:54c}) into Eq.~(\ref{eq:53}) and noting that $\langle \sigma_- \rangle_{\rm ss}=0$ [c.f. Eq.~(\ref{eq:qubitLiouvillian})], gives
\begin{widetext}
\begin{eqnarray}
{\rm Tr}_{\rm mc} [ q_{\rm p} \tilde{\mu} (\omega ) ] & = & -i \rho_{\rm q} \sum^{}_{\lambda}{}^{'} \frac{1}{i\omega - \lambda } \sum_{\rm p'} \bar{g}_{\rm p'q} \left[ \left\langle q_{\rm p} \hat{\Pi}_{\lambda_{\rm mc}} q_{\rm p'} \right\rangle_{\rm ss} {\rm Tr}_{\rm q} \left[ \sigma_- (- \lambda_{\rm mc} - \mathcal{L}_{\rm q} )^{-1} [ \sigma_x , \rho_{\rm q} ] \right] \right. \nonumber \\  
& & \left. + \left\langle [ q_{\rm p} \hat{\Pi}_{\lambda_{\rm mc}} , q_{\rm p'} ] \right\rangle_{\rm ss} \left( {\rm Tr}_{\rm q} [ \sigma_x (\lambda_{\rm mc} - \mathcal{L}_{\rm q} )^{-1} \sigma_- \rho_{\rm q} ] + {\rm Tr}_{\rm q} [ \sigma_- (- \lambda_{\rm mc} - \mathcal{L}_{\rm q} )^{-1} \rho_{\rm q} \sigma_x ] \right) \right] . \label{eq:55}
\end{eqnarray}
\end{widetext}

\subsection{Uncoupled Qubit Correlation Functions}
\label{app:uncoupledqubit}
Eq.~(\ref{eq:55}) gives the quantity that we need in terms of expectations of observables on the oscillator space and correlation functions on the qubit space. These correlation functions can be evaluated using the quantum regression theorem in the forms \cite{swain}
\begin{subequations}
\begin{eqnarray}
& & {\rm Tr}_{\rm q} \left[ \hat{A} (\lambda_{\rm mc} - \mathcal{L}_{\rm q})^{-1} \hat{B} \rho_{\rm q} \right] \nonumber \\
& = & \int^{+\infty}_0 dt \, e^{-\lambda_{\rm mc} t} \langle \hat{A}(t) \hat{B}(0) \rangle_{\rm ss} , \label{eq:QRT1} \\
& & {\rm Tr}_{\rm q} \left[ \hat{B} (\lambda_{\rm mc} - \mathcal{L}_{\rm q})^{-1} \rho_{\rm q} \hat{A} \right] \nonumber \\
& = & \int^{+\infty}_0 dt \, e^{-\lambda_{\rm mc} t} \langle \hat{A}(0) \hat{B}(t) \rangle_{\rm ss} , \label{eq:QRT2}
\end{eqnarray}
\end{subequations}
where $\hat{A}$ and $\hat{B}$ are arbitrary operators on the qubit space. The result is
\begin{eqnarray}
{\rm Tr}_{\rm mc} \left[ q_{\rm p} \tilde{\mu} (\omega ) \right] & = & -i \rho_{\rm q} \sum^{}_{\lambda}{}^{'} \frac{1}{i\omega - \lambda} \sum_{\rm p'} \bar{g}_{\rm p'q} \nonumber \\
& & \times \left[ r(\lambda_{\rm mc}) \langle q_{\rm p} \hat{\Pi}_{\lambda_{\rm mc}} q_{\rm p'} \rangle_{\rm ss} \right. \nonumber \\
& & \left. + t(\lambda_{\rm mc} ) \langle [ q_{\rm p} \hat{\Pi}_{\lambda_{\rm mc}} , q_{\rm p'} ] \rangle_{\rm ss} \right] , \nonumber \\
& & \label{eq:tracethingagain}
\end{eqnarray}
where $r(\lambda_{\rm mc})$ and $t(\lambda_{\rm mc})$ are uncoupled qubit correlation functions evaluated at the eigenvalues of the oscillator-space Liouvillian, see Eqs.~(\ref{eq:rapp}) and (\ref{eq:tapp}) of the Main Text. Given that the summation over the eigenvalues $\lambda$ in Eq.~(\ref{eq:tracethingagain}) explicitly excludes $\lambda_{\rm q}\neq 0$, we can recast it as a summation over the eigenvalues of the Liouvillian on the oscillator space, renormalised via their couplings to the qubit according to Eqs.~(\ref{eq:gammaminusp})-(\ref{eq:deltap}), $\lambda'_{\rm mc}$:
\begin{eqnarray}
{\rm Tr}_{\rm mc} \left[ q_{\rm p} \tilde{\mu} (\omega ) \right] & = & -i \rho_{\rm q} \sum^{}_{\lambda'_{\rm mc}}{}^{'} \frac{1}{i\omega - \lambda'_{\rm mc}} \sum_{\rm p'} \bar{g}_{\rm p'q} \nonumber \\
& & \times \left[ r(\lambda_{\rm mc}) \langle q_{\rm p} \hat{\Pi}_{\lambda_{\rm mc}} q_{\rm p'} \rangle_{\rm ss} \right. \nonumber \\
& & \left. + t(\lambda_{\rm mc} ) \langle [ q_{\rm p} \hat{\Pi}_{\lambda_{\rm mc}} , q_{\rm p'} ] \rangle_{\rm ss} \right] . \nonumber \\
& & \label{eq:traceappendix}
\end{eqnarray}
Now substituting Eq.~(\ref{eq:traceappendix}) into Eq.~(\ref{eq:Spomega}) leads to 
\begin{widetext}
\begin{eqnarray}
S_{\rm p}[\omega ] & = & {\rm Re} \, \sum_{\lambda'_{\rm mc}}{}^{'} \frac{1}{ i\omega - \lambda'_{\rm mc} } \sum_{\rm p'} \bar{g}_{\rm p'q} \left[ r(\lambda_{\rm mc}) \langle q_{\rm p} \hat{\Pi}_{\lambda_{\rm mc}} q_{\rm p'} \rangle_{\rm ss} + t(\lambda_{\rm mc}) \langle [ q_{\rm p} \hat{\Pi}_{\lambda_{\rm mc}} , q_{\rm p'} ] \rangle_{\rm ss} \right] \nonumber \\
& & \ \ \ \times \left( \tilde{g}_{\rm pq} {\rm Tr}_{\rm q} [ \sigma_+ \rho_{\rm q} ] - \bar{g}_{\rm pq} {\rm Tr}_{\rm q} [ \sigma_+ (i\omega - \mathcal{L}_{\rm q})^{-1} [ \sigma_x , \rho_{\rm q} ] ] \right) . \label{eq:secondhalf}
\end{eqnarray}
\end{widetext}
Using $\langle \sigma_+ \rangle_{\rm ss} = 0$ for the uncoupled qubit [c.f. Eq.~(\ref{eq:qubitLiouvillian})], we can drop the first term on the second line of Eq.~(\ref{eq:secondhalf}). Then we make the approximation $i\omega \rightarrow \lambda_{\rm mc}$ inside the qubit correlation functions, justified by the observation that the spectral peaks will appear near $\omega \sim {\rm Im} \, (\lambda_{\rm mc} )$. Using Eqs.~(\ref{eq:QRT1}) and (\ref{eq:QRT2}), the qubit correlation function in the second line of Eq.~(\ref{eq:secondhalf}) may be rewritten as
\begin{eqnarray}
& & {\rm Tr}_{\rm q} [ \sigma_+ (\lambda_{\rm mc} - \mathcal{L}_{\rm q})^{-1} [ \sigma_x , \rho_{\rm q} ] ] \nonumber \\
& = & \int^{+\infty}_0 dt \, e^{-\lambda_{\rm mc} t} \langle [\sigma_+(t) , \sigma_x (0)] \rangle . \label{eq:reqdqubitCF}
\end{eqnarray}
Substituting Eq.~(\ref{eq:reqdqubitCF}) into Eq.~(\ref{eq:secondhalf}) yields Eq.~(\ref{eq:oscillatorbit}) of the Main Text. 

The transformed uncoupled qubit correlation functions appearing in Eq.~(\ref{eq:oscillatorbit}) and defined in Eqs.~(\ref{eq:rapp}) and (\ref{eq:tapp}) are obtained from the Maxwell-Bloch equations corresponding to the Liouvillian of Eq.~(\ref{eq:qubitLiouvillian}) using the quantum regression theorem \cite{scully,woolley}. They are given by
\begin{subequations}
\begin{eqnarray}
r \left( \lambda_{\rm mc} \right) & = & \frac{ \gamma_\uparrow - \gamma_\downarrow }{\gamma_\downarrow + \gamma_\uparrow } \frac{1}{ \gamma_{\rm t}/2 + i \delta_{\rm d} - \lambda_{\rm mc} } , \label{eq:reval} \\
t \left( \lambda_{\rm mc} \right) & = & \frac{\gamma_\uparrow}{ \gamma_\downarrow + \gamma_\uparrow } \nonumber \\
& & \times \left\{ \frac{ \gamma_{\rm t}/2  - \mathrm{Re}(\lambda_{\rm mc}) + i [ \delta_{\rm d} + \mathrm{Im}(\lambda_{\rm mc}) ] }{ [ \gamma_{\rm t}/2  - \mathrm{Re}(\lambda_{\rm mc})  ]^2 + [ \delta_{\rm d} + \mathrm{Im}(\lambda_{\rm mc}) ]^2 } \right. \nonumber \\
& & \left. + \frac{ \gamma_{\rm t}/2 + \mathrm{Re}(\lambda_{\rm mc}) - i [ \delta_{\rm d} + \mathrm{Im}(\lambda_{\rm mc}) ] }{ [ \gamma_{\rm t}/2 + \mathrm{Re}(\lambda_{\rm mc})  ]^2 + [ \delta_{\rm d} + \mathrm{Im}(\lambda_{\rm mc}) ]^2 } \right\} , \nonumber \\ 
& & \label{eq:62b} \label{eq:teval} 
\end{eqnarray}
\end{subequations}
where we recall that $\delta_{\rm d}$ is the detuning between the qubit level splitting and the qubit drive frequency. Eq.~(\ref{eq:62b}) simplifies considerably if $\mathrm{Re}(\lambda_{\rm mc}) = 0$, to
\begin{equation}
t \left( \lambda_{\rm mc} \right) = \frac{\gamma_\uparrow}{ \gamma_\downarrow + \gamma_\uparrow } \frac{ \gamma_{\rm t} }{ (\gamma_{\rm t}/2)^2 + [ \delta_{\rm d} - \mathrm{Im}(\lambda_{\rm mc}) ]^2 } .
\end{equation}

\subsection{Oscillator-space Eigenvalues and Expectations: $16 g^2_{\rm mc} < ( \gamma_{\rm c,eff} - \gamma_{\rm m,eff})^2$}
\label{app:gmczero}
Next, we must evaluate the expectations and correlation functions appearing in Eq.~(\ref{eq:oscillatorbit}), in the presence of a direct electromechanical coupling satisfying (\ref{eq:weakcouplingassumption}). Under the condition (\ref{eq:weakcouplingassumption}) the imaginary parts of the eigenvalues of the oscillator Liouvillian are unchanged from the case in which the electrical circuit and mechanical modes are uncoupled. The eigenvalues may be written as ${\rm Im}\, (\lambda_{\rm mc}) = {\rm Im}\, (\lambda_{\rm m}) + {\rm Im}\, (\lambda_{\rm c})$ where ${\rm Im}\, (\lambda_{\rm p}) = k_{\rm p} \omega_{\rm p}$ with $k_{\rm p} = 0,\pm 1,\ldots$. We phenomenologically incorporate dissipation into a thermal environment by a simple modification of the eigenvalues, while leaving the eigenvectors unchanged, and justify this approximation using numerical calculations. 
Thus the oscillator-space eigenvalues, without and with the renormalisation due to coupling to the qubit, are
\begin{subequations}
\begin{eqnarray}
\lambda_{\rm mc} & = & \sum_{\rm p} \left[ ik_{\rm p} \omega_{\rm p} - (1-\delta_{k_{\rm p},0} ) \tilde{\gamma}_{\rm p}/2 \right] , \\
\lambda'_{\rm mc} & = & \sum_{\rm p} \left[ ik_{\rm p} (\omega_{\rm p} + \delta_{\rm p}) - (1-\delta_{k_{\rm p},0} ) \tilde{\gamma}_{\rm p,eff}/2 \right] , \nonumber \\
& & \label{eq:lambdaprimemc} 
\end{eqnarray}
\end{subequations}
respectively, where $\delta_{m,n}$ denotes the Kronecker delta, and $\tilde{\gamma}_{\rm p}$ and $\tilde{\gamma}_{\rm p,eff}$ are defined in Eq.~(\ref{eq:decaytilderates})~and~(\ref{eq:decaytilderateseff}). 

The eigenvectors corresponding to the eigenvalues having ${\rm Im}\, (\lambda_{\rm p}) = k_{\rm p} \omega_{\rm p}$ are then $| n_{\rm p} \rangle \langle n_{\rm p}+k_{\rm p} |$, where $|n_{\rm p} \rangle$ denotes a number state of oscillator ${\rm p}$. Since the oscillators are decoupled, the oscillator projection operators can be decoupled into two independent projection operators, $\hat{\Pi}_{\lambda_{\rm mc}} = \hat{\Pi}_{ k_{\rm p}} \otimes \hat{\Pi}_{ k_{\rm \bar{p}}}$. The action of the projector onto a subspace corresponding to the eigenvalue $k_{\rm p}$ of oscillator ${\rm p}$ is given by $\hat{\Pi}_{k_{\rm p}} \hat{A} = \sum^{+\infty}_{n_{\rm p}=0} | n_{\rm p} \rangle \langle n_{\rm p}+k_{\rm p} | \langle n_{\rm p} | \hat{A} | n_{\rm p}+k_{\rm p} \rangle$. It may be shown that $\hat{\Pi}_{\lambda_{\rm mc}} q_{\rm p} = \hat{\Pi}_{k_{\rm \bar{p}}} \left( a^\dagger_{\rm p} \delta_{k_{\rm p},-1} + a_{\rm p} \delta_{k_{\rm p},1} \right)$, and then it follows that
\begin{widetext} 
\begin{subequations}
\begin{eqnarray}
\langle [ q_{\rm p} \hat{\Pi}_{\lambda_{\rm mc} } , q_{\rm p} ] \rangle_{\rm ss} & = & \delta_{ k_{\rm \bar{p}},0 } ( \delta_{k_{\rm p},-1} - \delta_{k_{\rm p},+1} ) ( 1 + \langle (a^\dagger_{\rm p})^2 \rangle_{\rm ss} - \langle a^2_{\rm p} \rangle_{\rm ss} ) \rightarrow \delta_{ k_{\rm \bar{p}},0 } \left( \delta_{k_{\rm p},-1} - \delta_{k_{\rm p},+1} \right) , \label{eq:moment1} \\
\langle q_{\rm p} \hat{\Pi}_{\lambda_{\rm mc}} q_{\rm p} \rangle_{\rm ss} & = & ( \langle a^\dagger_{\rm p} a_{\rm p} \rangle_{\rm ss} + \langle a^2_{\rm p} \rangle_{\rm ss} ) \delta_{ k_{\rm \bar{p}},0 } \delta_{ k_{\rm p},+1 } + ( \langle a^\dagger_{\rm p} a_{\rm p} \rangle_{\rm ss} + \langle (a^\dagger_{\rm p})^2 \rangle_{\rm ss} + 1 ) \delta_{ k_{\rm \bar{p}},0 } \delta_{k_{\rm p},-1} \nonumber \\
& \rightarrow & \langle a^\dagger_{\rm p} a_{\rm p} \rangle_{\rm ss} \delta_{ k_{\rm p},+1 } \delta_{ k_{\rm \bar{p}},0 } + ( \langle a^\dagger_{\rm p} a_{\rm p} \rangle_{\rm ss} + 1 ) \delta_{k_{\rm p},-1} \delta_{ k_{\rm \bar{p}},0 } , \label{eq:moment3} \\
\langle q_{\rm p} \hat{\Pi}_{\lambda_{\rm mc}} q_{\rm \bar{p}} \rangle_{\rm ss} & = & \langle ( a_{\rm p} \delta_{k_{\rm p},-1} + a^\dagger_{\rm p} \delta_{k_{\rm p},+1} ) ( a^\dagger_{\rm \bar{p}} \delta_{ k_{\rm \bar{p}}, -1 } + a_{\rm \bar{p}} \delta_{k_{\rm \bar{p}},+1} ) \rangle \rightarrow 0 , \label{eq:moment4} \\
\langle [ q_{\rm p} \hat{\Pi}_{\lambda_{\rm mc}} , q_{\rm \bar{p}} ] \rangle_{\rm ss} & = & \langle a_{\rm p} ( a^\dagger_{\rm \bar{p}} - a_{\rm \bar{p}} ) \rangle_{\rm ss} \delta_{ k_{\rm p},-1 } \delta_{ k_{\rm \bar{p}},-1 } + \langle a_{\rm p} ( a_{\rm \bar{p}} - a^\dagger_{\rm \bar{p}} ) \rangle_{\rm ss} \delta_{ k_{\rm p},-1 } \delta_{ k_{\rm \bar{p}},+1 } \nonumber \\
& & + \langle a^\dagger_{\rm p} ( a^\dagger_{\rm \bar{p}} - a_{\rm \bar{p}} ) \rangle_{\rm ss} \delta_{ k_{\rm p},+1 } \delta_{ k_{\rm \bar{p}},-1 } + \langle a^\dagger_{\rm p} ( a_{\rm \bar{p}} - a^\dagger_{\rm \bar{p}} ) \rangle_{\rm ss} \delta_{ k_{\rm p},+1 } \delta_{ k_{\rm \bar{p}},+1 } \rightarrow 0 . \label{eq:moment5}
\end{eqnarray}
\end{subequations}
\end{widetext}
In evaluating Eqs.~(\ref{eq:moment1}) and (\ref{eq:moment3}) we have used the result $\langle \hat{\Pi}_{ k_{\rm \bar{p}} } \rangle = {\rm Tr}\, [ \hat{\Pi}_{k_{\rm \bar{p}}} \rho_{\rm ss}] = \delta_{k_{\rm \bar{p}},0}$. The limits in Eqs.~(\ref{eq:moment1})-(\ref{eq:moment5}) follow by assuming that phase-dependent oscillator moments and oscillator cross-correlations are negligible. This is expected to be the case for our proposed system with the anticipated experimental parameters.  

Using Eqs.~(\ref{eq:lambdaprimemc}), the specified limits of Eqs.~(\ref{eq:moment1})-(\ref{eq:moment5}), and Eqs.~(\ref{eq:reval})~and~(\ref{eq:teval}) in Eq.~(\ref{eq:oscillatorbit}) leads to the spectral contributions
\begin{widetext} 
\begin{eqnarray}
S_{\rm p}[\omega ] & = & \frac{\gamma_\downarrow - \gamma_\uparrow}{( \gamma_\downarrow + \gamma_\uparrow  )^2 } \sum_{\sigma = \pm 1} \frac{ 8 \bar{g}^2_{\rm pq} }{ [ (\gamma_{\rm t} + \tilde{\gamma}_{\rm p})^2 + 4 (\delta_{\rm d} + \sigma \omega_{\rm p})^2 ] [ \tilde{\gamma}^2_{\rm p,eff} + 4 ( \omega + \sigma ( \omega_{\rm p} + \delta_{\rm p} ) )^2 ] } \nonumber \\
& & \ \ \times \left\{ (\gamma_\downarrow - \gamma_\uparrow ) \tilde{\gamma}_{\rm p,eff} (\langle a^\dagger_{\rm p} a_{\rm p} \rangle + \delta_{\sigma,1} ) + 2\gamma_\downarrow \gamma_{\rm t} \frac{ \gamma_{\rm t} \tilde{\gamma}_{\rm p,eff} + 4 (\delta_{\rm d} + \sigma \omega_{\rm p} ) ( \omega + \sigma (\omega_{\rm p} + \delta_{\rm p}) ) }{ \gamma^2_{\rm t} + 4 (\delta_{\rm d} + \sigma \omega_{\rm p})^2 } \right\} . \nonumber \\
& & 
\end{eqnarray}
\end{widetext}
This contribution may be decomposed into an upper sideband $(\sigma = -1)$ and a lower sideband component $(\sigma = +1)$, which (after further simplifications) leads to Eqs.~(\ref{eq:sidebandsdecomposed})-(\ref{eq:lower}) of the Main Text.

\subsection{Oscillator-space Eigenvalues and Expectations: $16 g^2_{\rm mc} > ( \gamma_{\rm c,eff} - \gamma_{\rm m,eff})^2$}
\label{app:MotionalSideband2}

For a large direct electromechanical coupling, defined to mean that the condition (\ref{eq:weakcouplingassumption}) is \emph{not} satisfied, the presence of the coupling changes the imaginary part of the eigenvalues of the oscillator Liouvillian from the case in which the oscillators are uncoupled. Then the evaluation of Eq.~(\ref{eq:oscillatorbit}) is more conveniently performed in terms of the electromechanical \emph{normal-mode} operators, $\hat{a}_\pm = (\hat{a}_{\rm m} \pm \hat{a}_{\rm c} )/\sqrt{2}$. The Hamiltonian (\ref{eq:normalmodeHam}) may then be rewritten as
\begin{equation}
\hat{H}^{\rm mc}_{\rm S,RWA} = \hbar \sum_{ \sigma = \pm } (\omega_{\rm c} + \sigma g_{\rm mc} ) \hat{a}^\dagger_\sigma \hat{a}_\sigma .
\end{equation}

The Liouvillian eigenvalues corresponding to the Hamiltonian~(\ref{eq:normalmodeHam}), in the limit $16 g^2_{\rm mc} \gg ( \gamma_{\rm c,eff} - \gamma_{\rm m,eff})^2$, are
\begin{subequations}
\begin{eqnarray}
\lambda_{\rm mc} & = & \lambda_+ + \lambda_- \nonumber \\
& = & ik_+ ( \omega_{\rm c} + g_{\rm mc} ) + i k_- ( \omega_{\rm c} - g_{\rm mc}  ) \nonumber \\
& & - ( 1 - \delta_{k_+,0} \delta_{k_-,0} ) \sum_{\rm p} \gamma_{\rm p}/2, \\
\lambda'_{\rm mc} & = & i \sum_{ \sigma = \pm } k_\sigma \left[ \omega_{\rm c} + ( \delta_{\rm m} + \delta_{\rm c} )/2 \right. \nonumber \\
& & \left. + \sigma \sqrt{ ( \delta_{\rm m} - \delta_{\rm c} )^2/4 + g^2_{\rm mc} } \right] \nonumber \\
& & - ( 1 - \delta_{k_+,0} \delta_{k_-,0} ) \sum_{\rm p} ( \gamma_{\rm p} + \gamma^-_{\rm pe} - \gamma^+_{\rm pe} )/2 , \nonumber \\
& & \label{eq:eigenvaluesSum}
\end{eqnarray}
\end{subequations}
where $k_\pm = 0,\pm1,\ldots$, and $\lambda_{\rm mc}$ and $\lambda'_{\rm mc}$ correspond to oscillator-space eigenvalues without and with, respectively, the renormalisation due to the coupling to the qubit. The action of the corresponding projection operators on some operator $\hat{A}$ is given by
\begin{eqnarray}
\hat{\Pi}_{\lambda_{\rm mc}} \hat{A} & = & \sum^{+\infty}_{n_\pm = 0} ( | n_+ \rangle \langle n_+ + k_+ | \otimes | n_- \rangle \langle n_- + k_- | ) \nonumber \\
& & \times \langle n_+,n_- | \hat{A} | n_+ + k_+, n_- + k_- \rangle ,
\end{eqnarray}
where $| n_\pm \rangle$ are number states of the normal-mode oscillators. It may be shown that
\begin{eqnarray}
\sqrt{2} \hat{\Pi}_{\lambda_{\rm mc}} q_{\rm p} & = & a_+ \delta_{k_+,1} \delta_{k_-,0} + a^\dagger_+ \delta_{k_+,-1} \delta_{k_-,0}  \nonumber \\
& & \pm a_- \delta_{k_+,0} \delta_{k_-,1} \pm a^\dagger_- \delta_{k_+,0} \delta_{k_-,-1} , \nonumber \\
& & 
\end{eqnarray}
where the upper (lower) sign corresponds to ${\rm p} = {\rm m}$ (${\rm p} = {\rm c}$). The required moments for the evaluation of Eq.~(\ref{eq:oscillatorbit}) may then be calculated. 
Neglecting phase-dependent oscillator moments and oscillator cross-correlations, we find 
\begin{widetext}
\begin{subequations}
\begin{eqnarray}
2 \langle q_{\rm p} \hat{\Pi}_{\lambda_{\rm mc}} q_{\rm p} \rangle_{\rm ss} & = & \langle a^\dagger_{\rm p} a_{\rm p} \rangle_{\rm ss} ( \delta_{k_+,1} \delta_{k_-,0} + \delta_{k_+,0} \delta_{k_-,1} ) + \left( \langle a^\dagger_{\rm p} a_{\rm p} \rangle_{\rm ss} + 1 \right) ( \delta_{k_+,-1} \delta_{k_-,0} + \delta_{k_+,0} \delta_{k_-,-1} ) , \ \ \ \ \ \  \label{eq:lastone} \\
2 \langle q_{\rm p} \hat{\Pi}_{\lambda_{\rm mc}} q_{\rm \bar{p}} \rangle_{\rm ss} & = & \langle a^\dagger_{\rm p} a_{\rm p} \rangle_{\rm ss} ( \delta_{k_+,1} \delta_{k_-,0} - \delta_{k_+,0} \delta_{k_-,1} ) + \left( \langle a^\dagger_{\rm p} a_{\rm p} \rangle_{\rm ss} + 1 \right) ( \delta_{k_+,-1} \delta_{k_-,0} - \delta_{k_+,0} \delta_{k_-,-1} ) , \ \ \ \ \ \ \\
2 \langle [q_{\rm p} \hat{\Pi}_{\lambda_{\rm mc}} , q_{\rm p} ] \rangle_{\rm ss} & = & - \delta_{k_+,1} \delta_{k_-,0} + \delta_{k_+,-1} \delta_{k_-,0} - \delta_{k_+,0} \delta_{k_-,1} + \delta_{k_+,0} \delta_{k_-,-1} , \\
2 \langle [ q_{\rm p} \hat{\Pi}_{\lambda_{\rm mc}} , q_{\rm \bar{p}} \rangle_{\rm ss} & = & ( 1 + \langle a^\dagger_{\rm \bar{p}} a_{\rm \bar{p}} \rangle - \langle a^\dagger_{\rm p} a_{\rm p} \rangle ) \left( \delta_{k_+,0} \delta_{k_-,1} - \delta_{k_+,1} \delta_{k_-,0} \right) \nonumber \\
& & + ( 1 + \langle a^\dagger_{\rm p} a_{\rm p} \rangle - \langle a^\dagger_{\rm \bar{p}} a_{\rm \bar{p}} \rangle ) \left( \delta_{k_+,0} \delta_{k_-,1} - \delta_{k_+,1} \delta_{k_-,0} \right) . \label{eq:lastfour}
\end{eqnarray}
\end{subequations}
\end{widetext}
Substituting Eqs.~(\ref{eq:lastone})-(\ref{eq:lastfour}) into Eq.~(\ref{eq:oscillatorbit}), and summing over the eigenvalues of Eq.~(\ref{eq:eigenvaluesSum}) leads to the required motional sideband spectrum, with peaks split by the direct electromechanical coupling.

\end{document}